\title{Trust_Revision}

\documentclass[10pt,journal,compsoc]{IEEEtran}
\ifCLASSOPTIONcompsoc
  \usepackage[nocompress]{cite}
\else
  \usepackage{cite}
\fi
\ifCLASSINFOpdf
\else
\fi

\usepackage{cite,times,amsmath,epsfig,array}
\usepackage{mdwmath}
\usepackage{mdwtab}
\usepackage{eqparbox}
\usepackage{graphicx}
\usepackage{amsfonts}
\usepackage{epstopdf}
\usepackage{amsthm}
\usepackage{amssymb}
\usepackage{amsfonts}
\usepackage{algorithm}
\usepackage{algorithmic}

\begin{document}
%
\title{Clarifying Trust in Social Internet of Things}
%
%
%
%

\author{Zhiting~Lin
        and~Liang~Dong,~\IEEEmembership{Senior~Member,~IEEE}
\IEEEcompsocitemizethanks{\IEEEcompsocthanksitem Z.~T.~Lin is with School of Electronics and Information Engineering, Anhui University, Hefei, Anhui, 230601, China.\protect\\
E-mail: ztlin@ahu.edu.cn
\IEEEcompsocthanksitem L.~Dong is with Department of Electrical and Computer Engineering, Baylor University, Waco, Texas 76798, USA. \protect \\
E-mail: liang\_dong@baylor.edu}
}

%
%


\markboth{IEEE Transactions on Knowledge and Data Engineering}%
{Lin and Dong: Clarifying Trust in Social Internet of Things}
%



\IEEEtitleabstractindextext{%
\begin{abstract}
A social approach can be exploited for the Internet of Things (IoT) to manage a large number of connected objects.  These objects operate as autonomous agents to request and provide information and services to users.  Establishing trustworthy relationships among the objects greatly improves the effectiveness of node interaction in the social IoT and helps nodes overcome perceptions of uncertainty and risk.  However, there are limitations in the existing trust models.  In this paper, a comprehensive model of trust is proposed that is tailored to the social IoT.  The model includes ingredients such as trustor, trustee, goal, trustworthiness evaluation, decision, action, result, and context.  Building on this trust model, we clarify the concept of trust in the social IoT in five aspects such as (1) mutuality of trustor and trustee, (2) inferential transfer of trust, (3) transitivity of trust, (4) trustworthiness update, and (5) trustworthiness affected by dynamic environment.   With network connectivities that are from real-world social networks, a series of simulations are conducted to evaluate the performance of the social IoT operated with the proposed trust model.  An experimental IoT network is used to further validate the proposed trust model.
\end{abstract}

}

\maketitle

\IEEEdisplaynontitleabstractindextext

%
\IEEEpeerreviewmaketitle

\IEEEraisesectionheading{\section{Introduction}\label{sec:introduction}}

%
%
%
%
The Internet of Things (IoT) is evolving as a new generation of information network and service infrastructure, creating opportunities for more integration of the physical world into computer-based systems.  As a large number of objects are connected and the things get smart, it is indispensable for the interaction paradigm of the IoT to adopt a social approach~\cite{Atzori20123594,6710070,6802330}.  In a social IoT, the objects are capable of establishing social relationships with others and are allowed to have their own social networks.  The inter-object interactions occur on the objects' social network.  The social relationships among the users and owners are taken into account during the design phase of the IoT~\cite{6994231,7097037}.  The objects in the social IoT operate as autonomous agents to request and provide information and services to users while maintaining their individuality.

The social IoT has its advantages.  First, the structure of the social network can be shaped as required to guarantee network navigability and scalability.  As the number of objects connected to the network increases exponentially, the searching space becomes enormous~\cite{6994231,5951906}.  The heterogeneous nature and the large scale of contextual data make the IoT even more complicated~\cite{7045423}.  The social IoT can effectively perform the discovery of objects and services.  It navigates a social network of ``friendly'' objects instead of depending on typical Internet discovery tools which do not scale well.  Second, models designed to study social networks can be used to address issues of the social IoT.  These models are typically used for extensive networks with complicated and dynamic interconnections.  They can reveal how each object establishes social relationships and searches for information and services by crawling the networks.  Third, a level of trustworthiness can be established for leveraging the degree of interaction among objects that are friendly in the social IoT.  The social objects participate in a relationship only when there is enough trust.  The objects can effectively offer services to their owners by autonomously cooperating with other objects with which they have good relationships.

Trust relationships can exist between objects, making objects only respond to service requests from familiar nodes hence reducing exposure to malicious nodes~\cite{An2013799}.  On the other hand, when a task is transferred to a destination IoT agent for execution, the initiator of the task completely loses control of the task.  The task executor can easily manipulate the task code and attack the service requestor~\cite{XU2013107}.  Trust management helps the social IoT agents overcome perceptions of uncertainty and risk.

Most of the work to date focuses on narrow aspects of trust framework, trust evaluation, and trust transfer and inference~\cite{SurveyTrustIoT}.  There are still some misconceptions of trust in the social IoT and limitations in the existing trust models.  Developing a proper trust model is imperative for the design and implementation of the social IoT.  In this paper, we propose a comprehensive model of trust for the social IoT and discuss in detail the concepts of trust among objects.  Building on these concepts, we highlight five limitations of the existing trust models and clarify the distinctive features of trust in the social IoT.  The main contributions of this work are as follows.

\begin{enumerate}
\item A comprehensive model of trust is provided that is tailored to the social IoT.  In the social IoT, trust is more than a single concept such as trustworthiness.  It is described as a dynamic process rather than a static notion.  It has a relational construct of six basic ingredients, i.e., (1) the trustor, (2) the trustee, (3) the goal, (4) the evaluation of trustworthiness, (5) the decision and its subsequent action and result, and (6) the context.
\item Five limitations of current models of trust are discussed and the distinctive features of the trust model for the social IoT are clarified.  These five different aspects of the trust model include (1) mutuality of trustor and trustee, (2) inferential transfer of trust with analogous tasks, (3) transitivity of trust, (4) trustworthiness updated with delegation results, and (5) trustworthiness affected by dynamic environment.

 \item Because of features of the proposed model, it has the following merits: (1) providing protection of the trustee, (2) exploring information from characteristics and better using results of historical assignment, (3) offering two different scheme for transitivity of trust, (4) evaluating trust not only with positive factors but also with negative factors, and (5) adapting to dynamic environments.

\item A series of simulations and experiments are carried out to evaluate the performance of the social IoT which is operated with the proposed trust model.  The network connectivities of three real-world social networks, i.e., Facebook network, Google+ network, and Twitter, are used to construct the social IoT network.  Some characteristics of these real-world social network nodes are used as social IoT node characteristics.  An experimental IoT network is also used to validate the proposed trust model.
\end{enumerate}

The rest of the paper is organized as follows.  The related work on trust models is summarized in Section~\ref{sec:relatedwork}.  In Section~\ref{sec:model}, a general model of trust is proposed for the social IoT.  Six basic ingredients of the trust model are described.  In Section~\ref{sec:clarify}, the limitations of the existing trust models are listed.  Accordingly, we clarify the distinctive features of the trust model proposed for the social IoT.  In Section~\ref{sec:performace}, we evaluate the performance of the social IoT with methods based on the proposed trust model and compare it with common shortcomings of some models that are used in real-world systems~\cite{He2012,Zhan2013,Kantarci2014}.  Finally, conclusions and discussions are provided in Section~\ref{sec:conclusion}.

\section{Related Work}
\label{sec:relatedwork}
\subsection{Trust Models of IoT}
Recently, interest has piqued in the trust aspect of the IoT. There are some interesting models in this area. Deshpande et al.~developed a social network-based model to enable access controlled sharing of device capabilities in the IoT~\cite{7098685}. A prototype was implemented that uses public APIs to show the feasibility of the model. 
Daubert et al.~proposed a model that establishes a relation between information, privacy, and trust~\cite{7247581}.  The model balances between trust in the service provider and the need for privacy of individuals.

Many different aspects can be taken into consideration for calculating and modeling the trust, such as energy consumption, latency, and social relationships. Duan et al.~proposed an energy-aware trust derivation scheme, which aims to minimize energy consumption and latency of the network under the premise of security assurance~\cite{Duan2014}.  A trust derivation dilemma game was introduced into the trust derivation process to reduce the overhead of the network.  
Chen et al.~proposed an adaptive IoT trust protocol for Service-Oriented Architecture (SOA)-based IoT systems~\cite{6940301}.  For measuring social similarity and filtering trust feedback based on social similarity, they considered three social relationships, i.e., friendship, social contact, and community of interest. 
The effectiveness of the proposed adaptive IoT trust protocol was demonstrated through service composition application scenarios in SOA-based IoT environments when malicious nodes performed self-promoting, bad-mouthing, ballot-stuffing, and opportunistic service attacks.

Better understanding of the trust can facilitate the applications in IoT. Kantarci et al.~presented a Trustworthy Sensing for Crowd Management scheme for public safety~\cite{Kantarci2014}. Public safety authority can use sensor data of the smartphones if effective incentives exist for the users to provide the service.

\subsection{Trust Models of P2P Systems, Recommendation Systems, and Other Related Systems}

There is some excellent literature on trust models in related areas such as peer-to-peer (P2P) systems and recommendation systems. P2P is a suitable structure to realize the IoT. Related works in P2P systems can help us better understand the trust in IoT. Xiong and Liu introduced three basic trust parameters and two adaptive factors in computing trustworthiness of peers, namely, the feedback a peer receives from others, the total number of transactions a peer performs, the credibility of the feedback sources, the transaction context factor, and the community context factor~\cite{Xiong2004}. The feedback from those peers with higher credibility is weighted more than those with lower credibility.

Dewan et al.~investigated Reputation Systems for P2P networks, i.e., a more ambitious approach to protecting the P2P network without using any central component, and thereby harnessing the full benefits of the P2P network~\cite{Dewan2010}. The provider in their protocol is accountable for all past transactions, and cannot maliciously meddle with the transaction history by adding or deleting any recommendation. 

Nitti et al.~proposed a subjective model and an objective model to evaluate the object's trustworthiness that are derived from social networks and P2P technologies~\cite{6547148}.  
The subject approach has a slower transitory response. However, it is practically immune to certain malicious behaviors.

Recommendation systems are common in recent years which seek to predict the preference. Many aspects are considered in recommendation. Zhan et al.~introduced some shared character factors, such as credible feedback of digital contents, feedback weighting factor and user share similarity, and proposed a recommendation model~\cite{Zhan2013}. 
Chen et al.~proposed a generalized cross-domain collaborative filtering framework which integrates social network information seamlessly with cross-domain data~\cite{SIGKDD2013}.  
Recommendation using cross-domain data was carried out without decomposition.
Usually, the higher a user's expectation is prone to result in the lower a user's satisfaction. Therefore, Meng et al.~suggested that the evaluation vector should be adjusted so as to evaluate the server's service ability more accurately~\cite{Meng2016}.

Trust is one of the most important factor in the recommendation systems. Can and Bhargava defined three main trust metrics, i.e., reputation, service trust, and recommendation trust, to precisely measure trustworthiness~\cite{Can2013}. 
Guo et al.~suggested that not only the explicit but also the implicit influence of both rating and trust should be taken into consideration in a recommendation model~\cite{Guo2016}. 

The reputation or the trust can be subdivided. Fan et al.~defined two reputation values, i.e., recommended reputation value and recommending reputation value, for each peer to reflect the resource service behavior and trust recommending behavior, respectively~\cite{Fan2012}. Zhong et al.~proposed a trust model which distinguishes integrity trust from competence trust~\cite{Zhong2015}. 

He et al.~
identified the unique features of medical sensor networks and introduced relevant node behaviors, such as transmission rate and leaving time, into trust evaluation to detect malicious nodes~\cite{He2012}.
Das et al.~presented a dynamic trust computation model, named Secured Trust, to cope with the strategically altering behavior of malicious agents~\cite{Das2012}. 

These works have greatly enriched one's understanding of the challenges of the trust model.  However, the concept of trust in the IoT is still obscure, and there are some misunderstandings when frameworks or protocols are designed.   Fortunately, trust has been well studied in sociology~\cite{TrustTheory}.  Although we cannot directly apply those theories to the IoT, those theories can help us better understand trust in the IoT.

\section{A General Model of Trust in the Social Internet of Things}
\label{sec:model}

Before clarifying the characteristics of trust and discussing the limitations of current trust models, we provide a general model of trust in the social IoT.   As there is not yet a clear and prevailing notion of trust even in cognitive and social sciences, we layout a domain-specific definition of trust that is tailored to the social IoT.

\newtheorem*{mydef}{Definition}

\begin{mydef}
In the social IoT, trust is a process of the trustor, based on the evaluation and expectation of the trustee's competence and willingness, comprising the intention, deciding to delegate tasks to the trustee, and exploiting the outcome of the trustee's action for fulfilling a goal.  The trustor accepts the risk of becoming vulnerable by the act of entrusting the trustee in a certain context.  The evaluation of trustworthiness is mutual between the trustor and the trustee.  It depends on the task context and is affected by the behavior consequences and the environment uncertainty.
\end{mydef}

Trust in the social IoT is a relational construct of six basic ingredients: (1) the trustor, (2) the trustee, (3) the goal, (4) the evaluation of trustworthiness, (5) the decision and its subsequent action and result, and (6) the context.

\subsection{Trustor and Trustee}

Trustor, $X$, in the IoT is an intentional agent that has a goal, its own need, and attitudes toward other agents and their actions.  Based on its beliefs towards other agents and its cognition of the situation and the environment, the trustor can generate and delegate tasks and evaluate the results.
Trustee, $Y$, in the IoT is an agent equipped with devices that is capable of causing some effect as the outcome of its behavior.   In the case of social IoT, the trustee is also a cognitive agent.  Trustee $Y$ is another autonomous agent perceived by trustor $X$ and is beyond $X$'s direct control.  The behaviors of both the trustor and the  trustee have to be consistent with their trust relationship.

\subsection{Goal}

The trustor relies on the trustee's action to achieve a goal and/or to meet its own need.  With a goal, the trustor has the motivation to delegate tasks to the trustee and has the expectation of the result.  The expectation is positive if the trustee can produce the desired result which is favorable to achieving such a goal.  The expectation is negative if the result imposes frustration and threat against the goal.  The trustor intends to exploit the positive outcome of the trustee's action and is concerned to make decisions accordingly.

In the case of social IoT, trustor $X$ has no complete control over trustee $Y$.  Trustor $X$ takes risk by delegating tasks to trustee $Y$ and becomes vulnerable.  The trustor is vulnerable in terms of potential failure to achieve the goal.  The trustee may not perform the action or the action may not have the desired result.  There is uncertainty in the trustor's knowledge of the trustee.  In addition, when depending on the trustee for achieving the goal, the trustor is exposed to the potential damage inflicted by the trustee.

\subsection{Evaluation of Trustworthiness}

The trustor evaluates the trustee about its trustworthiness to do its share for achieving some goal.  Traditionally, trustworthiness is a property of the trustee perceived by the trustor.

In the social IoT, both the trustor and the trustee can be cognitive therefore the evaluation of trustworthiness is mutual.  Trustor $X$ evaluates trustee $Y$ and attributes to $Y$ an attitude and an expected action for achieving $X$'s goal.  At the same time, trustee $Y$ may evaluate trustor $X$ and attribute to $X$ a trustworthiness value in $Y$'s best interest, e.g., not to be maliciously exploited.
There are two types of trustworthiness evaluations in the social IoT, i.e., the pre-evaluation and the post-evaluation.  The trustor and the trustee pre-evaluate each other before the delegation action based on the context and past experiences.  The trustor tries to identify the best potential trustee and the trustee makes an effort to recognize malicious intents.  After the delegation action, the trustor and the trustee perform post-evaluations according to the results and the environment.  The evaluation is not only based on the success rate but also on the gain, the damage, the cost, and the environment.

\subsection{Decision, Action and Result}

In the social IoT, trust is a causal process that includes a decision, an action, and a result.  Trust is not merely an evaluation of or an attitude toward another agent.  It has its behavioral aspects in the decision making of the trustor and the subsequent course of action of the trustee~\cite{TrustTheory}.  The trustor evaluates the potential trustees, compares the expected outcomes, calculates its risks and costs, and creates an intension to delegate.  Upon its decision, the trustor delegates and relies on the trustee's action to produce the desired result.   If the trustee's behavior is predictable, the result of trust is the outcome of the expected action that can be exploited to fulfill the trustor's goal.  In practice, the result may deviate from what is expected and that will affect the relation between the trustor and the trustee.

Suppose that trustor $X$ can evaluate trustee $Y$ of performing task $\tau$.  The expected gain obtained by $X$ is $\hat{G}_{\mathrm{X} \leftarrow \mathrm{Y}}(\tau)$ if $Y$ accomplishes task $\tau$.  The expected damage suffered by $X$ is $\hat{D}_{\mathrm{X} \leftarrow \mathrm{Y}}(\tau)$ if $Y$ fails to do the task.  The expected cost of $X$ is $\hat{C}_{\mathrm{X} \leftarrow \mathrm{Y}}(\tau)$ regardless of $Y$'s success or failure.
The expected result of $Y$ executing task $\tau$ that can be exploited by $X$ is $\hat{R}_{\mathrm{X} \leftarrow \mathrm{Y}}(\tau)$, which is a function of $\hat{G}_{\mathrm{X} \leftarrow \mathrm{Y}}(\tau)$, $\hat{D}_{\mathrm{X} \leftarrow \mathrm{Y}}(\tau)$ and $\hat{C}_{\mathrm{X} \leftarrow \mathrm{Y}}(\tau)$. The expected gain, damage and cost can be expressed in terms of QoS/QoE parameters, such as delay, jitter, bandwidth, packet loss, procurement cost, reliability, efficiency, users' perspective of the overall value of the service provided, etc.  

Trustor $X$ has its goal $Goal_{\mathrm{X}}$.  If the expected result is aligned with the goal, e.g., $\hat{R}_{\mathrm{X} \leftarrow \mathrm{Y}}(\tau) \subseteq Goal_{\mathrm{X}}$, which means that the expected result is a subset of the goal, trustor $X$ delegates trustee $Y$ to do task $\tau$.  The outcome of $Y$'s action that can reach $X$ is the actual result $R_{\mathrm{X} \leftarrow \mathrm{Y}}(\tau)$.  The actual result may be different from the expected result.
Due to the lack of the expected outcomes and/or the addition of side effects, the actual result may not be a subset of the goal, i.e., $R_{\mathrm{X} \leftarrow \mathrm{Y}}(\tau) \nsubseteq Goal_{\mathrm{X}}$.
The expected gain $\hat{G}_{\mathrm{X} \leftarrow \mathrm{Y}}(\tau)$, damage $\hat{D}_{\mathrm{X} \leftarrow \mathrm{Y}}(\tau)$ and cost $\hat{C}_{\mathrm{X} \leftarrow \mathrm{Y}}(\tau)$ need to be modified accordingly.

\subsection{Context}

Trust is context dependent.  That is, the trustor trusts the trustee in a specific context about its behavior.  If the context is changed, the trustor's decision may be different.  The context consists of two components, i.e., the task type and the environment.
In the social IoT, trustor $X$ may trust trustee $Y$ for one task but not for another one.  The trustworthiness of an agent on performing one action can be different from performing another one.   The evaluation of trustworthiness needs to be applicable to the specific task.

The environment is an external condition.  In the trust process, there is perceived risk in the uncertainty of the actions of the autonomous agents as well as in the uncertainty of the environment.  Trustor $X$ evaluates the trustees and makes decision in a certain environment.   The environment affects trustor $X$'s evaluation process.   The environment also affects how trustee $Y$ acts, whether intentional or not, and how it generates the result.  The trustworthiness of $Y$ varies in different environments.  In the IoT, for example, the environment can be the supporting infrastructure or the external interference.  The trust process is situated on both task type and environment.

\begin{figure}[t]
\centering
\includegraphics[width=8.8cm]{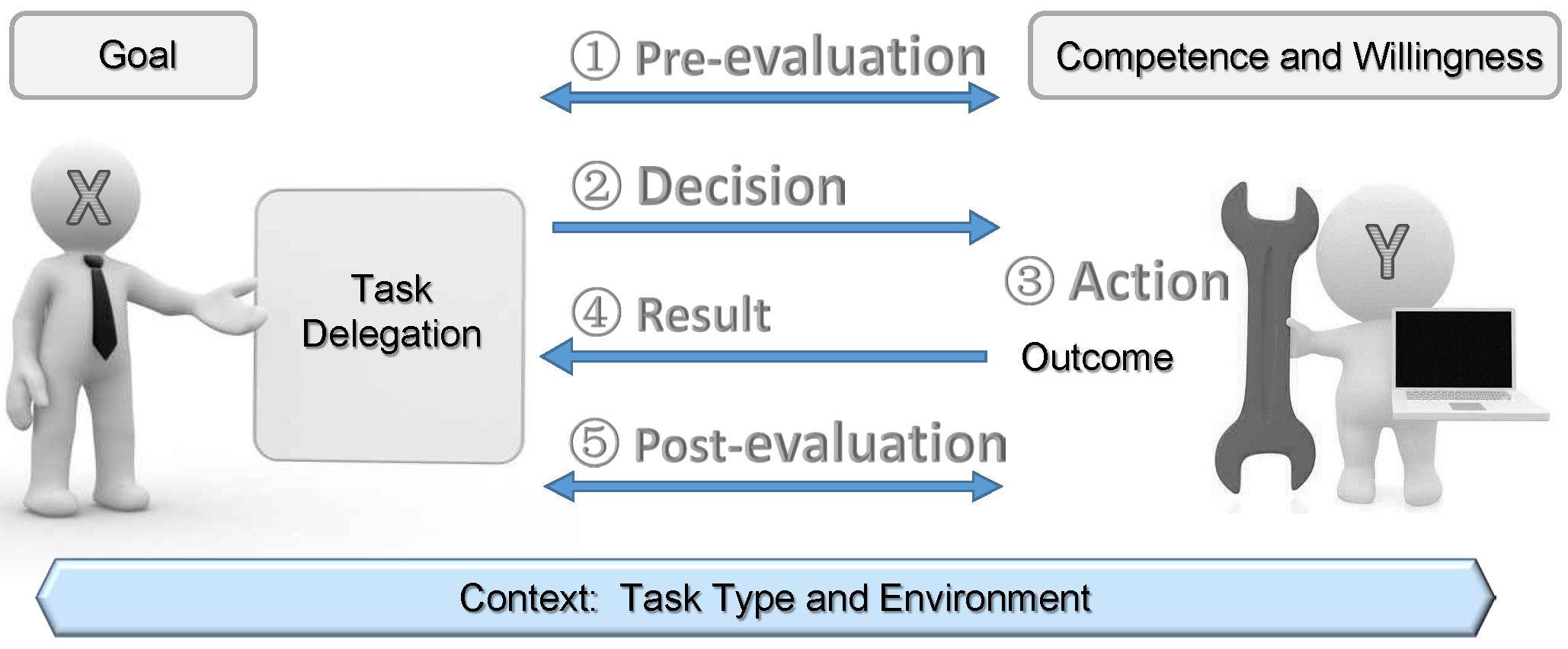}
\caption{Illustration of the ingredients and the process of trust.}
\label{fig:Fig1}
\end{figure}

The process of trust and all the ingredients are illustrated in Fig.~\ref{fig:Fig1}.  The notion of trust is more than a single value such as trustworthiness.  It is a dynamic process which involves the trustor, the trustee, and the circumstances rather than a static notion.   Trust contains not only a mental attitude, an evaluation and a decision but also an action full of unexpected risks.   In the following section, based on the described model of trust, we will highlight the limitations of some current approaches and clarify the distinctive features of trust in the social IoT.

\section{Clarification of Trust in the Social IoT}
\label{sec:clarify}

\subsection{Mutuality of Trustor and Trustee}

\newtheorem*{limit1}{Trust Model Limitation 1}

\begin{limit1}
In the trust process, the trustor performs a unilateral evaluation of the trustworthiness of the trustee.
\end{limit1}

A unilateral evaluation means that only the trustor performs evaluation on the trustee.  It delegates tasks to the trustee with expectations and risks.  This limitation of the existing trust models leads to a lack of protection of the trustee.  The trustee may want to evaluate the trustor's trustworthiness in order not to be maliciously exploited.  For instance, Alice (the trustor) intends to utilize Bob's (the trustee) camera installed at Bob's place.  Alice entrusts Bob with the ability of collecting information through his camera.  Meanwhile, Bob needs to make sure that Alice will not misuse the installed camera.

In the social IoT, both the trustor and the trustee are cognitive and the evaluation of trustworthiness is mutual so as to safeguard both sides' interests.  Before a decision of delegation, trustor $X$ pre-evaluates potential trustees and identifies the best candidate $Y$.  Candidate $Y$ performs a reverse evaluation towards $X$ based on $Y$'s own interest.  If $X$ passes the reverse evaluation, $Y$ becomes $X$'s trustee.  Moreover, trustor $X$ and trustee $Y$ perform mutual post-evaluations according to the action result and the environment, which will subsequently affect the pre-evaluations for the next delegation decision.  The mutual evaluation can be reflected in the following formulation of finding the trustee.
\begin{eqnarray}
\begin{array}{lll}
Y = &
\begin{split}
\mathop{\mathrm{argmax}}_{y}
\end{split}
& TW_{\mathrm{X} \leftarrow \mathrm{y}}(\tau) \\
& \mathrm{subject~to} & \widetilde{TW}_{\mathrm{y} \leftarrow \mathrm{X}}(\tau) \geq \theta_{y}(\tau)
\end{array}
\end{eqnarray}
where $TW_{\mathrm{X} \leftarrow \mathrm{y}}(\tau)$ denotes the trustworthiness of potential trustee $y$ perceived by trustor $X$ over task $\tau$,  $\widetilde{TW}_{\mathrm{y} \leftarrow \mathrm{X}}(\tau)$ denotes the reverse trustworthiness of trustor $X$ perceived by potential trustee $y$, and  $\theta_{y}(\tau)$ is a threshold set by $y$ for the reverse evaluation.  Only when the trustworthiness of the reverse evaluation is no less than $\theta_{y}(\tau)$, the trustee regards trustor $X$ as a trustworthy agent who will not maliciously exploit resources and accepts the delegation request.  To evaluate the trustor, the trustee can use its log files or usage pattern records to recognize how the trustor has used its resources. For example, if someone rents a server to provide illegal service, the server provider can detect it through its records.

\begin{figure}[t]
\centering
\includegraphics[width=8.8cm]{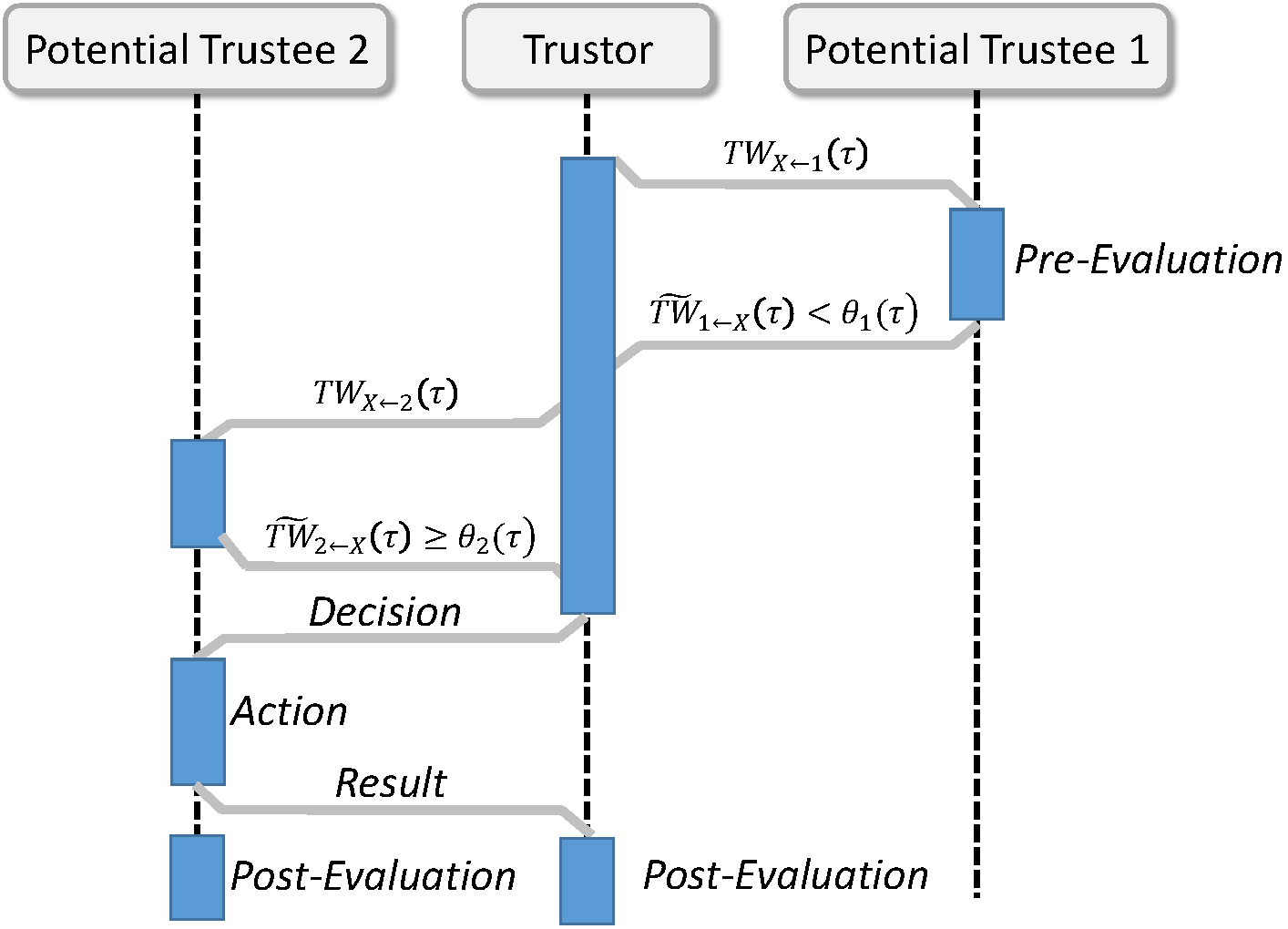}
\caption{Procedure of mutual evaluation on the trustee and the trustor.}
\label{fig:Fig2}
\end{figure}

Both pre-evaluation and post-evaluation in the social IoT are mutual.  The procedure of the mutual evaluation is illustrated in Fig.~\ref{fig:Fig2}.  Firstly, the trustor makes a pre-evaluation of all the potential trustees to delegate with task $\tau$.  It identifies potential trustee 1 with the highest trustworthiness $TW_{\mathrm{X} \leftarrow \mathrm{1}}(\tau)$.  Meanwhile, trustee 1 makes a reverse evaluation to compare the trustworthiness $\widetilde{TW}_{\mathrm{1} \leftarrow \mathrm{X}}(\tau)$ with its threshold $\theta_{1}(\tau)$.  Suppose that $\widetilde{TW}_{\mathrm{1} \leftarrow \mathrm{X}}(\tau) < \theta_{1}(\tau)$ and trustee 1 refuses to provide the service to trustor $X$.  Secondly, upon rejection, trustor $X$ chooses potential trustee 2 with the second best trustworthiness $TW_{\mathrm{X} \leftarrow \mathrm{2}}(\tau)$.  Suppose that, with reverse evaluation, trustee 2 agrees to provide the service.  Trustor $X$ makes decision to delegate trustee 2 to do task $\tau$ with expectations and risks.  Trustee 2 acts and produces the output, and trustor $X$ acquires the result.  Finally, both trustor $X$ and trustee 2 make post-evaluations based on the result and the environment.

\subsection{Inferential Transfer of Trust with Analogous Tasks}
\label{sec:AnalogousTasks}

\newtheorem*{limit2}{Trust Model Limitation 2}

\begin{limit2}
The trustor perceives the trustee's trustworthiness of performing a task as a single parameter that is unique to this specific task.  Therefore, when the trustor wants to delegate another task to the trustee, the trust based on the previous task cannot be transferred.
\end{limit2}

As trust is context dependent, conventionally, trustworthiness is unique to a specific task $\tau$.  $TW_{\mathrm{X} \leftarrow \mathrm{Y}}(\tau)$ indicates trustee $Y$'s chance of successfully performing task $\tau$ perceived by trustor $X$.  Numerous methods have been proposed to estimate the trustworthiness value, taking into account various factors such as competence, computation capability, willingness, and social contact~\cite{6547148}.  It is common for the trustor to use his experience to calculate the trustworthiness that is based on the trustee's past performance of task $\tau$.  However, restricting any trustworthiness to only one specific task is a limitation of the existing trust models.
In other words, when the trustor wants to delegate a new task to the trustee, the trust based on previous tasks cannot be used to infer the trustworthiness toward the new task.

Let us consider an example that Alice wants to check the real-time traffic of a certain route.  Bob claims that his smartphone can provide the related data.  Can Alice make a reasonable judgment based on the past experience that Bob's smartphone provided the GPS and image data? If the existing models are used, the answer is no. This is because GPS, image monitoring, and real-time traffic are deemed as three unrelated tasks in the existing models.  Although the real-time traffic monitoring task requires exactly the GPS and image information, it is considered as a new task.  Therefore, it has limits to treat each task as an inseparable entity.

In the trust model for social IoT, task $\tau$ includes multiple characteristics $\{a_{j}(\tau)\}_{j=1}^{J}$, where $a_{j}(\tau)$ denotes the $j$th characteristic of task $\tau$.  Based on this model, the trustworthiness of a new type of task $\tau'$ can be obtained from the existing trustworthiness values, even though trustor $X$ has not assigned $\tau'$ to trustee $Y$ previously.  One can infer trustworthiness $TW_{\mathrm{X} \leftarrow \mathrm{Y}}(\tau') $ with an inferring function $f$ and the existing trustworthiness value of $TW_{\mathrm{X} \leftarrow \mathrm{Y}}(\tau) $, if any characteristic $a_{i}(\tau')$ of the new task is included in the experienced task $\tau$.  That is, $\forall i, a_{i}(\tau'), \exists j, a_{j}(\tau)$, such that $a_{i}(\tau') = a_{j}(\tau)$, the trustworthiness with task $\tau'$ can be inferred as
\begin{equation}
TW_{\mathrm{X} \leftarrow \mathrm{Y}}(\tau') = f\left(TW_{\mathrm{X} \leftarrow \mathrm{Y}}(\tau)\right).
\end{equation}
If the characteristics $\{a_{i}(\tau')\}_{i=1}^{I}$ of new task $\tau'$ are included in multiple previously experienced tasks $\{\tau_{k}\}_{k=1}^{K}$, the trustworthiness with task $\tau'$ can be inferred from trustworthiness of all these tasks as
\begin{equation}
TW_{\mathrm{X} \leftarrow \mathrm{Y}}(\tau') = f\left(TW_{\mathrm{X} \leftarrow \mathrm{Y}}(\tau_{1}), \ldots, TW_{\mathrm{X} \leftarrow \mathrm{Y}}(\tau_{K})\right).
\end{equation}

\begin{figure}[t]
\centering
\includegraphics[width=7cm]{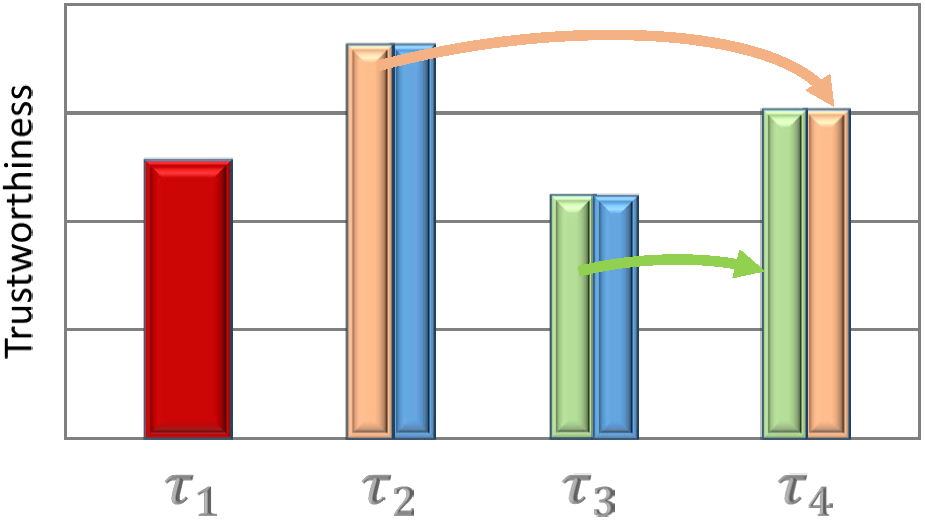}
\caption{An example of inferring trustworthiness of tasks.  A task may include multiple characteristics.}
\label{fig:Fig3}
\end{figure}

Fig.~\ref{fig:Fig3} illustrates an example of how to infer the trustworthiness of a new task from the trustworthiness of previously experienced tasks.  Different colors in the figure indicate different characteristics.  Assume that task $\tau_{1}$ includes only one characteristic which is indicated by red color. Meanwhile, tasks $\tau_{2}$ and $\tau_{3}$ consist of two characteristics, respectively.  Hence, two different colors in each of them.  Trustor $X$ requests trustee $Y$ to do a new type of task, $\tau_{4}$, whose characteristics are included in $\tau_{2}$ and $\tau_{3}$.  Although the trustor does not have direct experience of delegating $\tau_{4}$ to the trustee, he can still infer the trustworthiness.

Given that different characteristics play different roles in a task, each characteristic needs to be weighted to reflect its importance in the task.  Suppose that the $i$th characteristic in new task $\tau'$ is weighted as $w_{i}(\tau')$.  Therefore, an implementation of the inferring function $f$ can be given as
\begin{eqnarray}
TW_{\mathrm{X} \leftarrow \mathrm{Y}}(\tau') &=& \sum_{i} w_{i}(\tau') \frac{\sum_{k}w_{j}(\tau_{k}) TW_{\mathrm{X} \leftarrow \mathrm{Y}}(\tau_{k})}{\sum_{k}w_{j}(\tau_{k})} \nonumber \\
&& \mathrm{where}~~~ a_{i}(\tau') = a_{j}(\tau_{k}).
\end{eqnarray}
Here, it is assumed that multiple previous tasks $\{\tau_{k}\}$ have characteristics $a_{j}(\tau_{k})$ that are the same as characteristic $a_{i}(\tau')$ in the new task.
$w_{j}(\tau_{k})$ is a weight factor that represents the importance of the $j$th characteristic in task $\tau_{k}$.  As characteristic $a_{j}(\tau_{k})$ is the same as characteristic $a_{i}(\tau')$,
$\sum_{k}w_{j}(\tau_{k}) TW_{\mathrm{X} \leftarrow \mathrm{Y}}(\tau_{k}) / \sum_{k}w_{j}(\tau_{k})$ denotes the weighted average of the existing trustworthiness toward characteristic $a_{i}(\tau')$.  It is an estimation of how trustee $Y$ would do with characteristic $a_{i}(\tau')$ in task $\tau'$.  Eventually, $TW_{\mathrm{X} \leftarrow \mathrm{Y}}(\tau')$ is obtained as a weighted sum of these estimations of the characteristics that compose task $\tau'$.

The proposed trust model is a characteristic-based model, which has a broad range of applications.  It is suitable for the applications with a task that requires a set of characteristics.  For example, the real-time traffic monitoring task requires the GPS, the image data, and the velocity information. An agent can be entrusted of a task if it can undertake all characteristics of the task.  The trustworthiness of different characteristics can be evaluated through different previous tasks.    

\subsection{Transitivity of Trust}
\label{sec:Transitivity}

\newtheorem*{limit3}{Trust Model Limitation 3}

\begin{limit3}
If Alice trusts Bob and Bob trusts Carlos, Alice can trust Carlos without restrictions.
\end{limit3}

Transitivity of trust in social IoT means that if agent $X$, who requests the service, and agent $Y$, who provides it, are not linked by a direct social relationship, the trustworthiness value can be transferred via the intermediate social nodes~\cite{6547148}.  In existing models, the trustworthiness value of nonadjacent nodes $X$ and $Y$ is computed as
\begin{equation}
TW_{\mathrm{X} \leftarrow \mathrm{Y}} = \prod_{a,b \in \mathcal{P}^{XY}} TW_{\mathrm{a} \leftarrow \mathrm{b}} \label{eq:trans_model}
\end{equation}
where $\mathcal{P}^{XY}$ represents the sequence of nodes which constitute the selected path from node $X$ to node $Y$.  The model is a good simplification.  Nevertheless, this model does not distinguish task types.   Neither does it differentiate the recommendation from the task execution. Trust is simply transited as long as there is positive trustworthiness value between any two sequential nodes in path $\mathcal{P}^{XY}$.  In other words, trust is transited without any restriction in the existing models.
However, it has been demonstrated that in real life trust is not always transitive but depends on the particular service requested~\cite{Christianson1997}.  Therefore, to remedy this limitation, it is better to model the transitivity with restrictions that are based on the context.

Transitivity of the proposed trust model is obtained as a function $g$ of trustor $X$, trustee $Y$, path of intermediate nodes $\mathcal{P}^{XY}$, and the type of task $\tau$ and its recommendation $R_{\tau}$.  That is
\begin{equation}
TW_{\mathrm{X} \leftarrow \mathrm{Y}}(\tau) = g(X, Y, \mathcal{P}^{XY}, \tau, R_{\tau}).
\end{equation}
The intermediate nodes provide recommendation rather than service, as $R_{\tau}$ denotes the recommendation for task $\tau$.   The type of the task is emphasized in the model and the trust transitivity is discussed in the following two situations.

\begin{figure}[t]
\centering
\includegraphics[width=8.2cm]{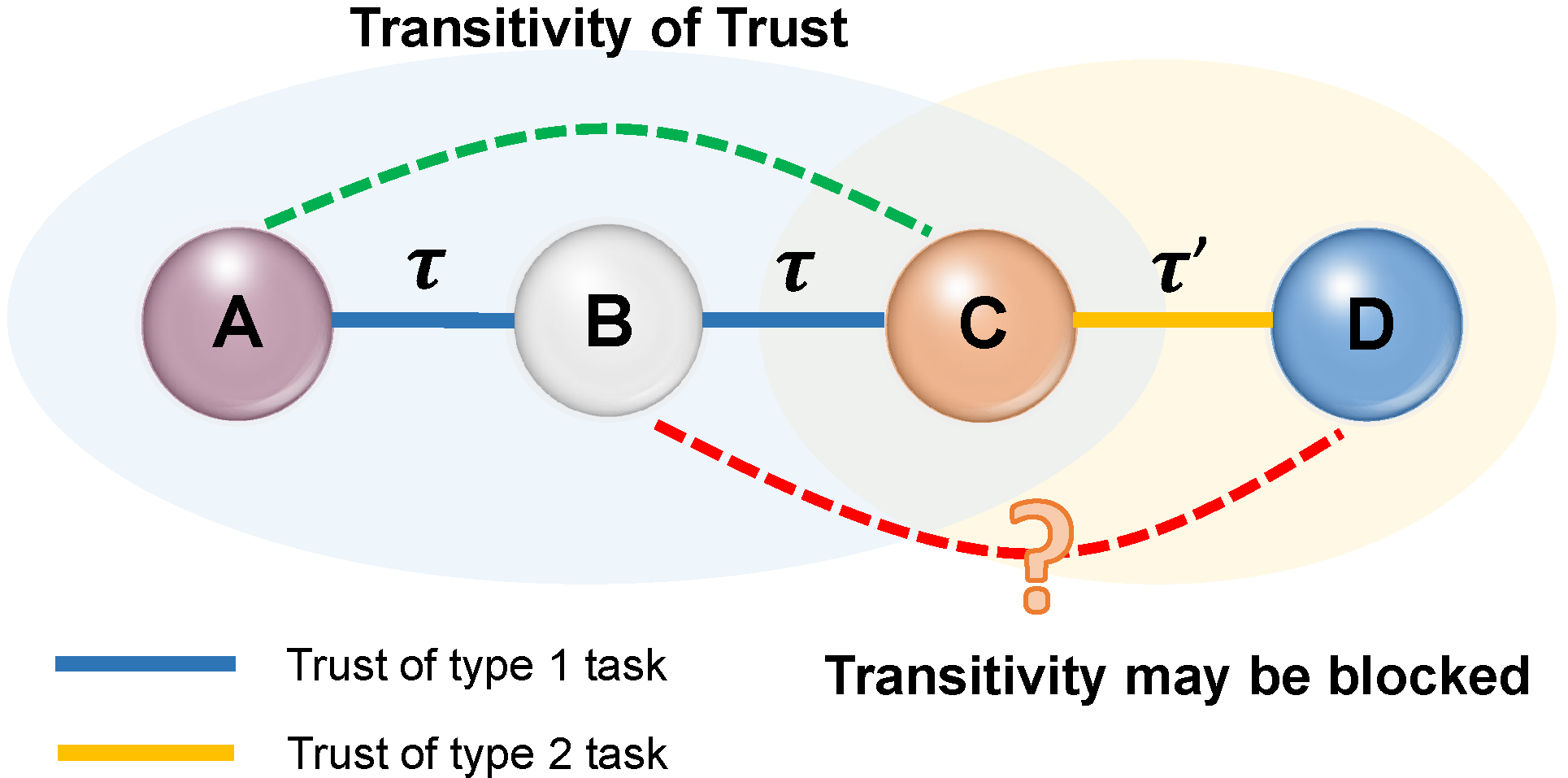}
\caption{Transitivity of trust with respect to tasks of the same type.}
\label{fig:Fig4}
\end{figure}

In the first situation, the type of the task does not change over the transitivity path.  If Alice trusts Bob and Bob trusts Carlos with the same task type, how can Alice infer trustworthiness toward Carlos of this task type?  As illustrated in Fig.~\ref{fig:Fig4}, trust of Alice (A) toward Bob (B) is based on task of type 1 and that of Bob (B) toward Carlos (C) is based on the task of the same type.  The transitivity is allowed.  Trust can be transited when A regards B as a competent intermediate node, i.e., $TW_{\mathrm{A \leftarrow B}}(R_{\tau}) \geq \omega_{1}$, and B regards C as a suitable trustee, i.e., $TW_{\mathrm{B \leftarrow C}}(\tau) \geq \omega_{2}$.  Here, $\omega_{1}$ and $\omega_{2}$ are preset trustworthiness with relatively high values.  The transition of trust is given by
\begin{IEEEeqnarray}{lll} \label{eq:transitivity}
TW_{\mathrm{A \leftarrow C}}(\tau) &&~=~ TW_{\mathrm{A \leftarrow B}}(R_{\tau}) TW_{\mathrm{B \leftarrow C}}(\tau) \nonumber \\
&&~~~~~~+ \left(1 - TW_{\mathrm{A \leftarrow B}} (R_{\tau}) \right) \left(1-TW_{\mathrm{B \leftarrow C}}(\tau) \right)\nonumber\\
&&~=~1-TW_{\mathrm{A \leftarrow B}}(R_{\tau})-TW_{\mathrm{B \leftarrow C}}(\tau) \nonumber \\
&&~~~~~~+ 2 \cdot TW_{\mathrm{A \leftarrow B}} (R_{\tau})TW_{\mathrm{B \leftarrow C}}(\tau) 
\end{IEEEeqnarray}
It is included in \eqref{eq:transitivity} a part of the transitivity of trust, i.e., $ (1 - TW_{\mathrm{A \leftarrow B}} (R_{\tau})) (1-TW_{\mathrm{B \leftarrow C}}(\tau))$, which is neglected in the existing model \eqref{eq:trans_model}.  It represents a mistrust toward the intermediate node multiplied by the incorrect judgment of the intermediate node toward its predecessor. 
If the task types are different along the path, the transitivity of trustworthiness may be blocked as in the case of $B \leftarrow C \leftarrow D$ in Fig.~\ref{fig:Fig4}.

In the second situation, the task types along the transitivity paths are different.  However, these tasks have some common characteristics.  With the concepts in Section~\ref{sec:AnalogousTasks}, two methods are proposed here for calculating the transitivity of trust.

\begin{figure}[t]
\centering
\includegraphics[width=8.8cm]{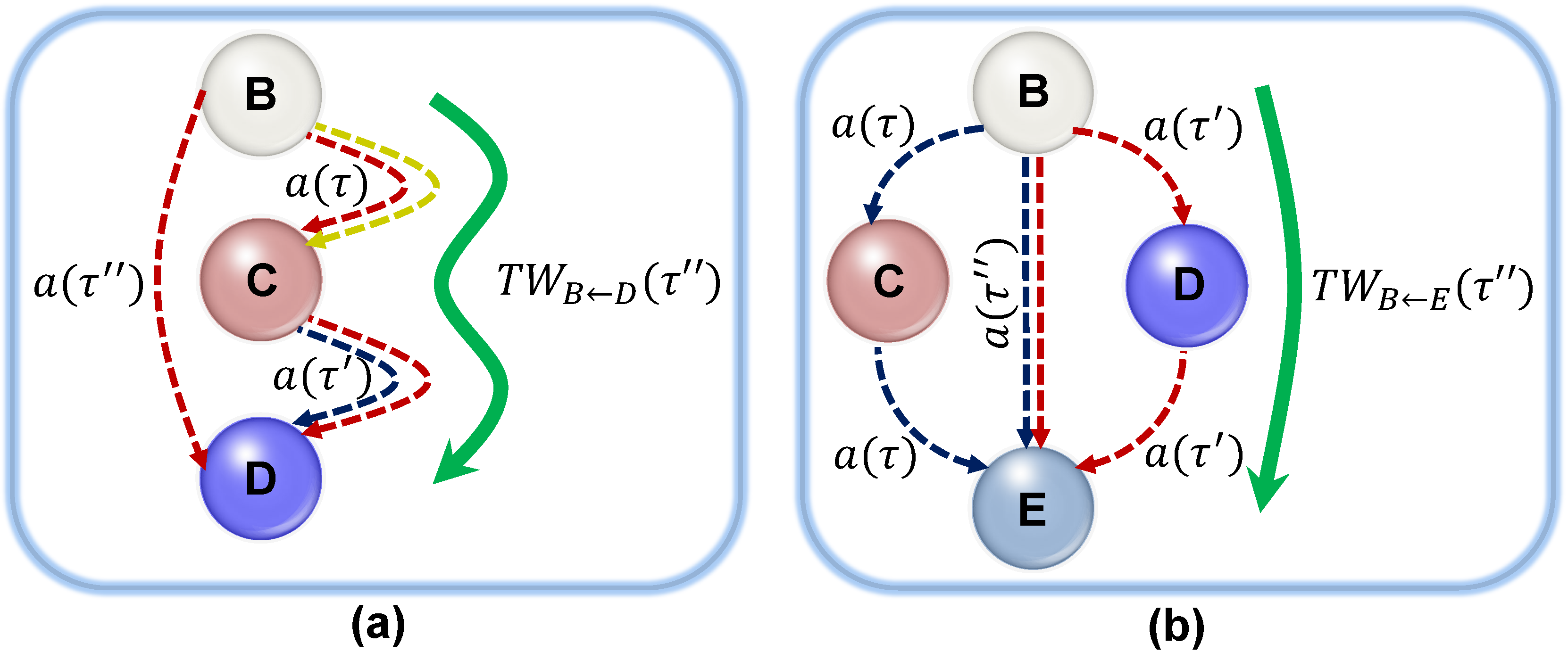}
\caption{Transitivity of trust with respect to tasks that have multiple characteristics. (a) Conservative Transitivity.  (b) Aggressive Transitivity.}
\label{fig:Fig5}
\end{figure}

(1) Conservative Transitivity.

Trustworthiness can be inferred from the intermediate social nodes
only if all the characteristics of the new task are included in the experienced tasks of the intermediate
nodes.  Take Fig.~\ref{fig:Fig5}(a) as an example.   If Bob (B) trusts Carlos (C) with task $\tau$ and Carlos (C) trusts Dale (D) with task $\tau'$, Bob (B) can infer trustworthiness toward Dale (D) with task $\tau''$ when
\begin{equation}
\{a(\tau'')\} \subseteq \{a(\tau)\} \cap \{a(\tau')\}.
\end{equation}
The trustworthiness of task $\tau''$ can be inferred as
\begin{eqnarray}
TW_{\mathrm{B \leftarrow C}}(R_{\tau''}) &=& f(TW_{\mathrm{B \leftarrow C}}(R_{\tau})) \\
TW_{\mathrm{C \leftarrow D}}(\tau'') &=& f(TW_{\mathrm{C \leftarrow D}}(\tau')).
\end{eqnarray}
The transitivity of trust is given by
\begin{eqnarray}
\lefteqn{TW_{\mathrm{B \leftarrow D}}(\tau'') = TW_{\mathrm{B \leftarrow C}}(R_{\tau''}) TW_{\mathrm{C\leftarrow D}}(\tau'') } \nonumber \\
&~~~~~~~~& + (1-TW_{\mathrm{B \leftarrow C}}(R_{\tau''})) (1- TW_{\mathrm{C\leftarrow D}}(\tau''))\nonumber \\
\lefteqn{~~when~~TW_{\mathrm{B \leftarrow C}}(R_{\tau''})>\omega_1~~and~~TW_{\mathrm{C\leftarrow D}}(\tau'')\omega_2.}
\end{eqnarray}

(2) Aggressive Transitivity.

Trustworthiness can be inferred from the intermediate social nodes
if any of the characteristics of the new task is included in the experienced tasks of the intermediate
nodes along a path and all the characteristics are included in the experienced tasks of the trustee.  It means that the assessment of different characteristics of a particular task can be done along different paths.  Take Fig.~\ref{fig:Fig5}(b) as an example.  Bob (B) trusts Carlos (C) and Carlos (C) trusts Evan (E) with the same task $\tau$.  Bob (B) trusts Dale (D) and Dale (D) trusts Evan (E) with the same task $\tau'$.  Bob (B) can infer trustworthiness toward Evan (E) with task $\tau''$ when
\begin{equation}
\{a(\tau'')\} \subseteq \{a(\tau)\} \cup \{a(\tau')\}.
\end{equation}
Suppose that characteristics $\{a_{1}\}$ pass through path $B \leftarrow C \leftarrow E$ and characteristics $\{a_{2}\}$ through path $B \leftarrow D \leftarrow E$.
The trustworthiness of characteristics $\{a_{1}\}$ and $\{a_{2}\}$ of task $\tau''$ can be inferred as
\begin{eqnarray}
TW_{\mathrm{B \leftarrow C}}(R_{a_{1}(\tau'')}) &=& f(TW_{\mathrm{B \leftarrow C}}(R_{\tau})) \\
TW_{\mathrm{C \leftarrow E}}(a_{1}(\tau'')) &=&  f(TW_{\mathrm{C \leftarrow E}}(\tau)) \\
TW_{\mathrm{B \leftarrow D}}(R_{a_{2}(\tau'')}) &=& f(TW_{\mathrm{B \leftarrow D}}(R_{\tau'})) \\
TW_{\mathrm{D \leftarrow E}}(a_{2}(\tau'')) &=&  f(TW_{\mathrm{D \leftarrow E}}(\tau')).
\end{eqnarray}
Furthermore, the trustworthiness of the characteristics need to be combined to establish the trustworthiness of task $\tau''$.  That is
\begin{eqnarray}
TW_{\mathrm{B \leftarrow E}}(\tau'') &=& w_{1}(\tau'') TW_{\mathrm{B \leftarrow E}}(a_{1}(\tau'')) \nonumber \\
&& + w_{2}(\tau'') TW_{\mathrm{B \leftarrow E}}(a_{2}(\tau''))
\end{eqnarray}
where $TW_{\mathrm{B \leftarrow E}}(a_{1}(\tau''))$ and $TW_{\mathrm{B \leftarrow E}}(a_{2}(\tau''))$  can be obtained with the transitivity equations similar to \eqref{eq:transitivity}.
The searching complexity and communication overhead are larger of the Aggressive Transitivity method than those of the Conservative Transitivity method.  Nevertheless, more potential trustees can be found with the Aggressive Transitive method.

\subsection{Trustworthiness Updated with Delegation Results}
\label{sec:UpdatingTrustworthiness}

\newtheorem*{limit4}{Trust Model Limitation 4}

\begin{limit4}
It is not well modeled how the trustworthiness is updated with the results of previous task delegations.
\end{limit4}

There are two kinds of trustworthiness evaluation in the social IoT.  One is pre-evaluation and the other is post-evaluation. In most of the existing trust models, the trustor modifies the trustworthiness toward the trustee after task delegation.  However, it is not clear how the results that are fed back affect the evaluation in the existing models.

The proposed trust model emphasizes on the various aspects of the delegation results.  It is reasonable to evaluate the trustworthiness with other factors in addition to the success rate.  For example, the energy of a social IoT node may be limited because it is powered by a battery or a renewable energy source.  The energy consumption of previous tasks greatly impacts the willingness of this node to undertake any more similar tasks.  The factors of delegation results can be classified into positive factor \emph{gain} and negative factors \emph{damage} and \emph{cost}~\cite{Falcone2001}.  Therefore, the normalized post-evaluation of the trustworthiness is given by
\begin{IEEEeqnarray}{lll} \label{eq:post_evaluation}
TW_{\mathrm{X \leftarrow Y}}(\tau) &~=~& \mathrm{N}\left[\hat{S}_{\mathrm{X \leftarrow Y}}(\tau) \hat{G}_{\mathrm{X \leftarrow Y}}(\tau) - (1 - \hat{S}_{\mathrm{X \leftarrow Y}}(\tau)) \right. \nonumber \\
&& \left. \cdot \hat{D}_{\mathrm{X \leftarrow Y}}(\tau) - \hat{C}_{\mathrm{X \leftarrow Y}}(\tau) \right]
\end{IEEEeqnarray}
where $\hat{S}_{\mathrm{X \leftarrow Y}}(\tau)$ denotes the expected success rate of trustee $Y$ completing task $\tau$.  $\hat{G}_{\mathrm{X \leftarrow Y}}(\tau)$ is the expected gain to trustor $X$ by assigning task $\tau$ to $Y$ and $Y$ completes the task.  $\hat{D}_{\mathrm{X \leftarrow Y}}(\tau)$ is the expected damage infringed to $X$ by assigning task $\tau$ to $Y$ but $Y$ fails the task.  $\hat{C}_{\mathrm{X \leftarrow Y}}(\tau)$ is the expected cost of $X$ delegating task $\tau$ to $Y$ regardless of the outcomes.  $\mathrm{N}[\cdot]$ denotes the normalization operator that brings the trustworthiness value to a specified range, e.g., $[0,1]$ or $[-1,1]$.  These expected results are updated with the actual success rate $S_{\mathrm{X \leftarrow Y}}(\tau)$, gain $G_{\mathrm{X \leftarrow Y}}(\tau)$, damage $D_{\mathrm{X \leftarrow Y}}(\tau)$ and cost $C_{\mathrm{X \leftarrow Y}}(\tau)$ of the current task delegation.  That is
\begin{eqnarray}
\hat{S}_{\mathrm{X \leftarrow Y}}(\tau) &=& \beta \hat{S}_{\mathrm{X \leftarrow Y}}'(\tau) + (1-\beta) S_{\mathrm{X \leftarrow Y}}(\tau) \label{eq:update_S}\\
\hat{G}_{\mathrm{X \leftarrow Y}}(\tau) &=& \beta \hat{G}_{\mathrm{X \leftarrow Y}}'(\tau) + (1-\beta) G_{\mathrm{X \leftarrow Y}}(\tau) \\
\hat{D}_{\mathrm{X \leftarrow Y}}(\tau) &=& \beta \hat{D}_{\mathrm{X \leftarrow Y}}'(\tau) + (1-\beta) D_{\mathrm{X \leftarrow Y}}(\tau) \\
\hat{C}_{\mathrm{X \leftarrow Y}}(\tau) &=& \beta \hat{C}_{\mathrm{X \leftarrow Y}}'(\tau) + (1-\beta) C_{\mathrm{X \leftarrow Y}}(\tau)  \label{eq:update_C}
\end{eqnarray}
where $\hat{\bullet}'$ denotes the historical expected value and $\beta$ is the forgetting factor.  It should be noted that $\beta$ can be set to different values in the above four updating equations.

The proposed model clarifies how to update the trustworthiness according to the delegation results.
Metrics of social relationships, such as friendship and community-interest, can be used to calculate the initial  values of the trustworthiness.  For example, socially cooperative nodes in the same community tend to provide high performance.   Since social relationship enhancement can be considered as a kind of benefit or gain, it can be modeled and included in the gain factor $\hat{G}(\tau)$ in the proposed model.   There are some excellent existing works that detail how social relationships can be involved in calculating the trustworthiness~\cite{6994231,7097037}.

If the trustworthiness value is large, the trustor is likely to get positive net profit by delegating the task to the trustee.  Without the normalization in \eqref{eq:post_evaluation}, some assignments may lead to positive net profits while others may result in negative net profits.   A rational task assignment is the one that can bring the most expected profit.  Therefore, the best candidate to delegate task $\tau$ satisfies
\begin{eqnarray}
Y =&
\begin{split}
\mathop{\mathrm{argmax}}_{y}
\end{split} &
\hat{S}_{\mathrm{X \leftarrow y}}(\tau) \hat{G}_{\mathrm{X \leftarrow y}}(\tau) - (1 - \hat{S}_{\mathrm{X \leftarrow y}}(\tau)) \nonumber \\
&& \cdot \hat{D}_{\mathrm{X \leftarrow y}}(\tau) - \hat{C}_{\mathrm{X \leftarrow y}}(\tau).
\end{eqnarray}

With the proposed model of updating the trustworthiness with the delegation results, it is easy to include the trustor itself as one of the candidates to perform the task.  In the social IoT, an agent  trusting others to accomplish a task does not necessarily mean that the requester cannot do the job by himself~\cite{TrustTheory}.  Although the agent has resource and capability to accomplish the task, he trusts and delegates the task to others if there is more net profit.  That is, trustor $X$ assigns task $\tau$ to trustee $Y$ rather than do it himself if
\begin{IEEEeqnarray}{c}
\hat{S}_{\mathrm{X\leftarrow Y}}(\tau) \hat{G}_{\mathrm{X\leftarrow Y}}(\tau) - (1 - \hat{S}_{\mathrm{X\leftarrow Y}}(\tau)) \hat{D}_{\mathrm{X\leftarrow Y}}(\tau) - \hat{C}_{\mathrm{X\leftarrow Y}}(\tau)   \nonumber \\
 >   \nonumber \\
\hat{S}_{\mathrm{X\leftarrow X}}(\tau) \hat{G}_{\mathrm{X\leftarrow X}}(\tau) - (1 - \hat{S}_{\mathrm{X\leftarrow X}}(\tau)) \hat{D}_{\mathrm{X\leftarrow X}}(\tau) - \hat{C}_{\mathrm{X\leftarrow X}}(\tau). \nonumber \\
\end{IEEEeqnarray}

When an agent of the social IoT is entrusted with a task request, he also has two options. He can either complete the task or recommend and delegate to other agents to do it.  The decision is based on which option can bring more benefits to himself.   As the trustee evaluates the trustor, its own goal
is to obtain more gain and suffer less damage and cost.   Usually, the trustee's gain includes earnings and reputation improvement. The damage and cost include equipment amortization, energy consumption, bandwidth occupation, etc. The reverse evaluation can use a similar equation as \eqref{eq:post_evaluation}.

\subsection{Trustworthiness Affected by Dynamic Environment}
\label{sec:TrustworthinessContext}

\newtheorem*{limit5}{Trust Model Limitation 5}
\begin{limit5}
The update of trustworthiness depends entirely on the task delegation results without considering the dynamic of the social IoT environment.
\end{limit5}

Since trust is context dependent, the update of trustworthiness should be based on the delegation results as well as on the context~\cite{TrustTheory}.  The context is an important factor that affects trust, because it specifies the situation in which trust resides~\cite{SurveyTrustIoT}.  As we stated before, the context consists of the task types and the environment.  The same environment that is safe to one agent/task may be hostile to another agent/task.  Given the dynamic nature of a social IoT, the environment often changes and the threat exposure may vary considerably over time~\cite{Køien2011}.  Furthermore, people behave differently in different situations such that the devices associated with humans may change their behaviors in different environments.

Dynamic environment means that the external condition changes considerably.  It is imperative to adjust trust assessment and trustworthiness assignment accordingly.  Chen \emph{et al.}~\cite{7097037} included a time factor in the trustworthiness update to adapt to the environment variation.   Trustor $X$ utilizes its latest experience with trustee $Y$ and the previous trustworthiness values to update the trustworthiness.  However, it is not sufficient to model the effect of the dynamic environment.   For instance, it is more difficult for any agent to accomplish a task in a hostile environment than an amicable one.
Hostile environment means that the external condition is unsuitable and harmful for accomplishing the current task.
Amicable environment means that the external condition is suitable and helpful for accomplishing the current task.
It is reasonable for the trustor to update the trustworthiness of the trustee with an extra reward if it accomplishes the task in a hostile environment.

The instantaneous environment, i.e., the current external condition, of trustor $X$ is modeled as $E_{\mathrm{X}}$ and the instantaneous environment of trustee $Y$ is modeled as $E_{\mathrm{Y}}$.  Suppose that $\mathcal{I}$ is the set of the intermediate nodes that connect $X$ and $Y$.  The instantaneous environments of the intermediate nodes are modeled as $\{E_{i}\}, i \in \mathcal{I}$. In order to stabilize the trustworthiness updates, we introduce a function $r(\cdot)$ to ``remove'' the environment influence on the actual success rate, gain, damage, and cost perceived by the trustor.  Consequently, the update functions \eqref{eq:update_S}--\eqref{eq:update_C} are modified as
\begin{eqnarray}
\hat{S}_{\mathrm{X \leftarrow Y}}(\tau) &=& \beta \hat{S}_{\mathrm{X \leftarrow Y}}'(\tau) + (1-\beta) \label{eq:updateS} \nonumber \\
&& r(E_{\mathrm{X}}, E_{\mathrm{Y}}, \{E_{i}\},  S_{\mathrm{X \leftarrow Y}}(\tau)) \\
\hat{G}_{\mathrm{X \leftarrow Y}}(\tau) &=& \beta \hat{G}_{\mathrm{X \leftarrow Y}}'(\tau) + (1-\beta) \nonumber \\
&& r(E_{\mathrm{X}}, E_{\mathrm{Y}}, \{E_{i}\},  G_{\mathrm{X \leftarrow Y}}(\tau)) \\
\hat{D}_{\mathrm{X \leftarrow Y}}(\tau) &=& \beta \hat{D}_{\mathrm{X \leftarrow Y}}'(\tau) + (1-\beta) \nonumber \\
&& r(E_{\mathrm{X}}, E_{\mathrm{Y}}, \{E_{i}\},  D_{\mathrm{X \leftarrow Y}}(\tau)) \\
\hat{C}_{\mathrm{X \leftarrow Y}}(\tau) &=& \beta \hat{C}_{\mathrm{X \leftarrow Y}}'(\tau) + (1-\beta)  \nonumber \\
&& r(E_{\mathrm{X}}, E_{\mathrm{Y}}, \{E_{i}\},  C_{\mathrm{X \leftarrow Y}}(\tau)).
\end{eqnarray}

Let us use a coarse example to illustrate how function $r(\cdot)$ works.  Suppose that the environment indicators $E_{\mathrm{X}}$, $E_{\mathrm{Y}}$, and $\{E_{i}\}$ take real positive values in $(0,1]$ where a large number represents amicable environment and a small number represents hostile environment.  Function $r$ can be defined as
\begin{equation}
 r(E_{\mathrm{X}}, E_{\mathrm{Y}}, \{E_{i}\},  S_{\mathrm{X \leftarrow Y}}(\tau)) = \frac{S_{\mathrm{X \leftarrow Y}}(\tau)}{\min [E_{\mathrm{X}}, E_{\mathrm{Y}}, \{E_{i}\} ]}. \label{eq:functionr}
\end{equation}
In this way, accomplishing a task in a hostile environment has extra credit on trustworthiness.  Here, the smallest value of the instantaneous environment is used according to Cannikin Law (Wooden Bucket Theory) because the worst environment has the dominant influence.

\section{Performance Evaluation}
\label{sec:performace}

\subsection{Real-world Social Networks for Network Connectivity in Social IoT Simulations}

We adopt three real-world social networks for network connectivity and perform simulations to evaluate the clarified trust models for the social IoT.  The network connectivity cases of the actual Facebook network, Google+ network, and Twitter network are used as the connectivity case of the simulated IoT.

The Facebook data were collected from survey participants of Facebook users. The dataset includes node features (user profiles) and circles (user's friend lists with direct connections).
The Google+ data were collected from users who manually shared their circles using the ``share circle" feature.   The Twitter data were crawled from public sources. The dataset also includes node features and circles~\cite{Mcauley:2014:DSC:2582178.2556612}.

Due to the complexity of these social networks, we use sub-networks extracted from these real-world networks for simulations of the social IoT.  The connectivity characteristics of the sub-networks are presented in Table~\ref{tab:socialnetworks}.
The degree of a node is the number of edges that it connects to. The average degree of all the nodes gives an overall indication of the degree of connectivity of the network.   The diameter is the largest number of steps of the shortest paths between nodes in the network.  The average path length is the average number of steps of the shortest paths of all pairs of nodes.

\begin{table}[b]
\centering
\caption{Connectivity characteristics of the three sub-networks of social networks.}
\label{tab:socialnetworks}
    \begin{tabular}{ | l | c | c | c |}
    \hline
     & Facebook & Google+ & Twitter \\ \hline \hline
    Number of Nodes & 347 &	358 & 244 \\ \hline
    Number of Edges & 5038 & 4178 & 2478 \\ \hline
    Average Degree & 29.04 & 23.34 & 20.31  \\ \hline
    Diameter & 11 & 12 & 8  \\ \hline
    Average Path Length & 3.75 & 3.9 & 2.96  \\ \hline
    Average Clustering Coefficient & 0.49 & 0.39 & 0.27 \\ \hline
    Modularity & 0.46 & 0.45 & 0.38 \\ \hline
    Number of Communities & 29 & 22 & 16 \\ \hline
    \end{tabular}
\end{table}

The clustering coefficient is given by the ratio of the number of edges to the maximum possible number of edges in a node's direct neighborhood.   Clustering coefficients indicate how nodes are embedded in their neighborhood.
The clustering coefficient, along with the average path length, usually indicate a ``small-world" effect.   The average clustering coefficient is the mean value of the clustering coefficients of all the network nodes~\cite{ICWSM09154}.
The modularity values shown in Table~\ref{tab:socialnetworks} reveal that these three sub-networks are loosely concentrated in modules (groups of densely connected nodes)~\cite{newman2006modularity}.  The number of communities (groups) of each sub-network is also shown in Table~\ref{tab:socialnetworks}~\cite{1742-5468-2008-10-P10008}.

With each sub-network, we randomly select about 40\% of the nodes as trustors and about 40\% of the nodes as trustees for a social IoT.  The social networks simulation platform is used to verify the clarified trust models 1, 3, 4, and 5 in Sections~\ref{sec:mutuality}, \ref{sec:transitivity},  \ref{sec:delegation}, and \ref{sec:dynamic_environment}, respectively.

\subsection{Experiment Setup}
\label{sec:experiment_setup}

\begin{figure}[t]
\centering
\includegraphics[width=8.4cm]{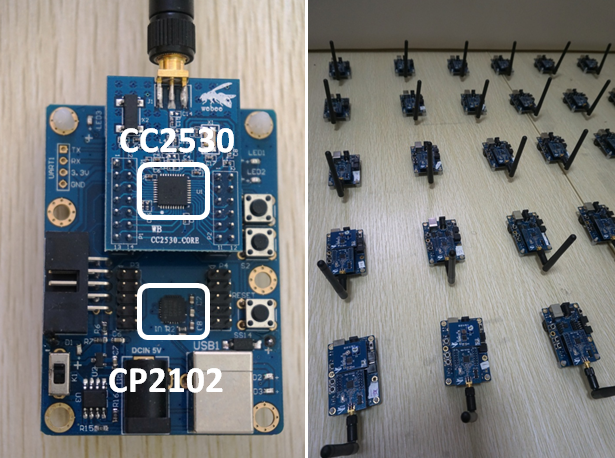}
\caption{Node devices of the experimental IoT network.}
\label{fig:experimentsetup}
\end{figure}

Besides the real-world social networks platform, we carry out experiments in an IoT network to test the proposed trust models.  Each node device in the IoT network is installed with the Texas Instruments' Z-Stark (version 2.5.0).   The Z-Stack includes five layers, i.e., the ZigBee Device Objects layer, the Application Framework, the Application Support Sublayer, the ZigBee network layer, and the ZMAC layer~\cite{zstack}.   These layers support the devices to construct a ZigBee network. 

The node devices used in our experiments have a size of $3.6 \times 2.7$ cm.  It contains a CC2530 chip.   The CC2530 is a system-on-chip (SoC) solution for IEEE 802.15.4, Zigbee, and Radio Frequency for Consumer Electronics (RF4CE) applications.  It enables robust network nodes that are built with very low total bill-of-material costs. The CC2530 combines the excellent performance of an RF transceiver with an industry-standard enhanced 8051 Microcontroller Unit (MCU).   All the I/Os of the CC2530 chip are available through 2.54 pin interfaces and can be used to extend the function.   Optical sensors are attached to the main boards by these 2.54 pin interfaces and used in Section~\ref{sec:dynamic_environment}.   Each device uses a 2.4 GHz omnidirectional antenna.   Its reliable transmission distance is up to 250 meters, and the automatic reconnection distance is up to 110 meters.   Fig.~\ref{fig:experimentsetup} shows the node devices of the experimental IoT network.

We establish an experimental IoT network that contains five node groups.  Each group includes two trustors, two honest trustees, and two dishonest trustees.  Besides, there is a coordinator device that is configured to start the IEEE 802.15.4 network.  It is the first device on the network.  The coordinator scans the RF environment, chooses a channel and a network identifier, and starts the network.  At the end of each experiment, the coordinator collects the data and sends them back to the host computer through a CP2102 chip for further analysis.  The CP2102 is a highly integrated USB serial port conversion module.  

The experiment platform is used to verify the clarified trust models 2, 4, and 5 in Sections~\ref{sec:inferential_transfer}, \ref{sec:delegation}, and \ref{sec:dynamic_environment}, respectively.

\subsection{Mutuality in Trust Model}
\label{sec:mutuality}

In order to evaluate the trust model with mutuality of trustor and trustee, a trustor is assigned a trustworthiness value by its potential trustee through reverse evaluation. The trustworthiness value represents how the trustor would use the trustee’s resources legitimately and responsively. It is based on the trustee’s previous experience with the trustor. The better the previous usage (less abusive), the greater the trustworthiness value through the reverse evaluation.  Any potential trustee $y$ holds a threshold $\theta_{y}(\tau)$ for task $\tau$.  It only accepts delegation requests from the trustors whose trustworthiness values are greater than $\theta_{y}(\tau)$.

In the simulation, we assign each trustor a trustworthiness value which is a random number in $[0, 1]$.  If this value is high, the trustor uses the trustee's resources responsively with a high probability.  If this value is low, the trustor behaves maliciously and uses the trustee's resources abusively with a high probability.
We assume that the reverse evaluation is performed based on the statistics of the trustor's previous responsive or abusive uses of the trustee's resources.  For task $\tau$, potential trustee $y$ sets a threshold $\theta_{y}(\tau)$ that is equal to 0, 0.3, or 0.6.  When $\theta_{y}(\tau) = 0$, it means that the trustee accepts delegation requests from any trustor.  This is equivalent to the case of unilateral evaluation that only the trustor evaluates the trustee.

\begin{figure}[t]
\centering
\includegraphics[width=0.98\linewidth]{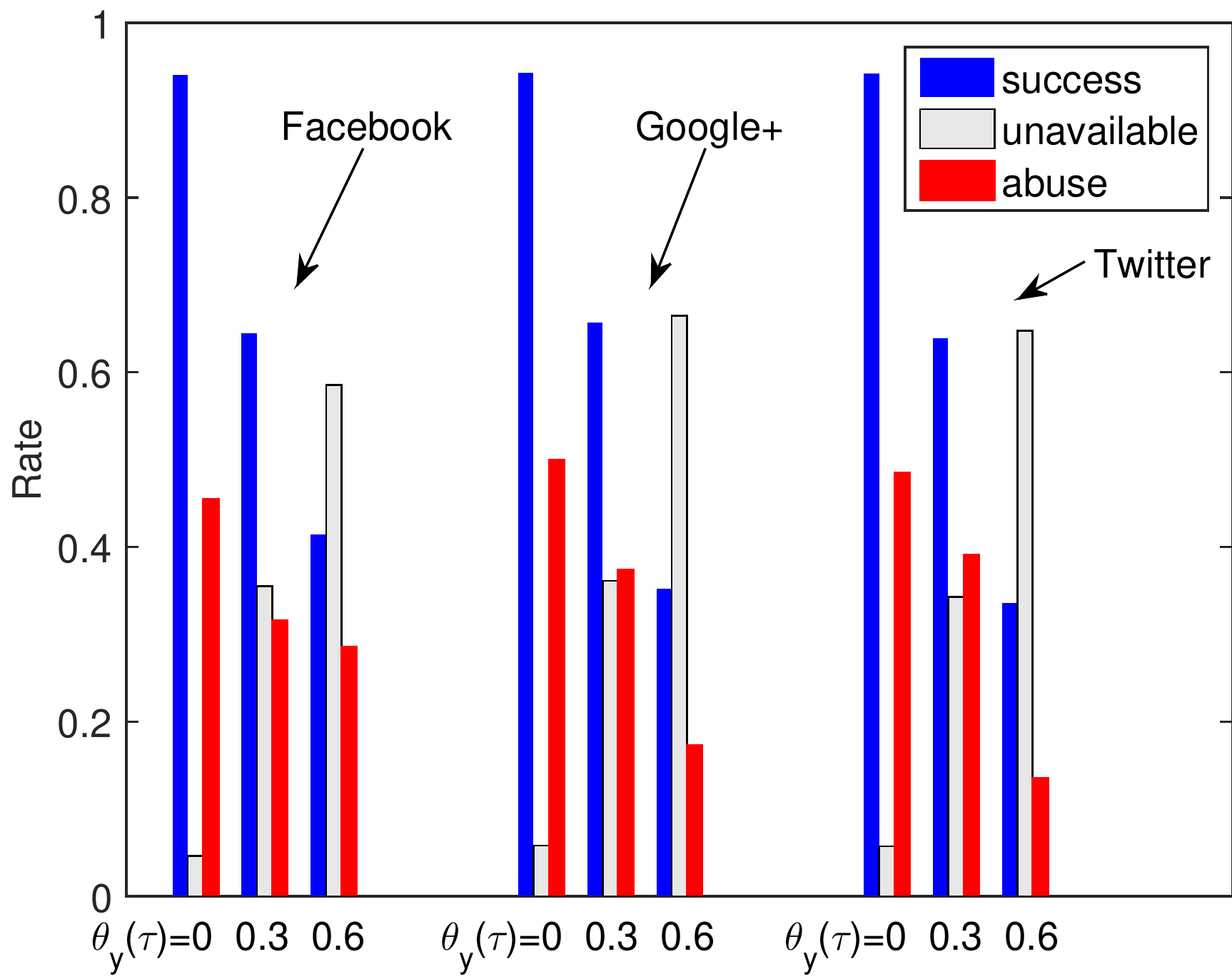}
\caption{Comparison of success rates, unavailable rates, and abuse rates of task delegations with different threshold value $\theta_{y}(\tau)$ in the reverse evaluations.}
\label{fig:section52}
\end{figure}

Fig.~\ref{fig:section52} shows the success rate, unavailable rate, and abuse rate of one delegating task $\tau$ to another in the sub-networks of Facebook, Google+, and Twitter.   The success rate is the ratio of the number of successful task delegations to the total number of delegation requests.   The unavailable rate is the ratio of the number of unanswered requests to the total number of requests.   Some trustors may not find any trustee to accept task $\tau$ because of the low trustworthiness values in the reverse evaluations.  With task delegations, the abuse rate is the ratio of the number of abusive uses to the number of all uses of the trustees' resources.

It is revealed in the figure that, if the trustees do not perform the reverse evaluation and accept all  requests, i.e., $\theta_{y}(\tau) = 0$, the abuse rates are more than 0.4 in the three networks.  As the threshold $\theta_{y}(\tau)$ increases, the unavailable rates increase and the abuse rates decrease across all networks. Some vicious nodes cannot obtain services when the trustees execute rigorous assessments in the reverse evaluations. There are some differences in the results among these three networks.   This is because the structures of Facebook, Google+ and Twitter are different.   For example, the average degree is 29.04 for Facebook, 23.34 for Google+, and 20.31 for Twitter.

\subsection{Inferential Transfer of Trust in Trust Model}
\label{sec:inferential_transfer}

The proposed trust model traces down to the multiple characteristics within a task.  It allows the trustworthiness of a new task to be inferred from the analogous tasks.  Once a malicious trustee behaves poorly on one task, it affects subsequent evaluations of this trustee with other types of tasks that contain the same characteristics.  

To test the effectiveness of the proposed trust model, we use the experiment platform of the IoT network described in Section~\ref{sec:experiment_setup}.  Each of the trustors requests a task that contains two characteristics.   The characteristics are also included in different previous tasks.  Dishonest trustees have performed maliciously with a particular characteristic on the previous task.  And, the trustors make a judgment that the dishonest trustees are not as competent as the honest trustees with that characteristic.  

\begin{figure}[t]
\centering
\includegraphics[width=0.98\linewidth]{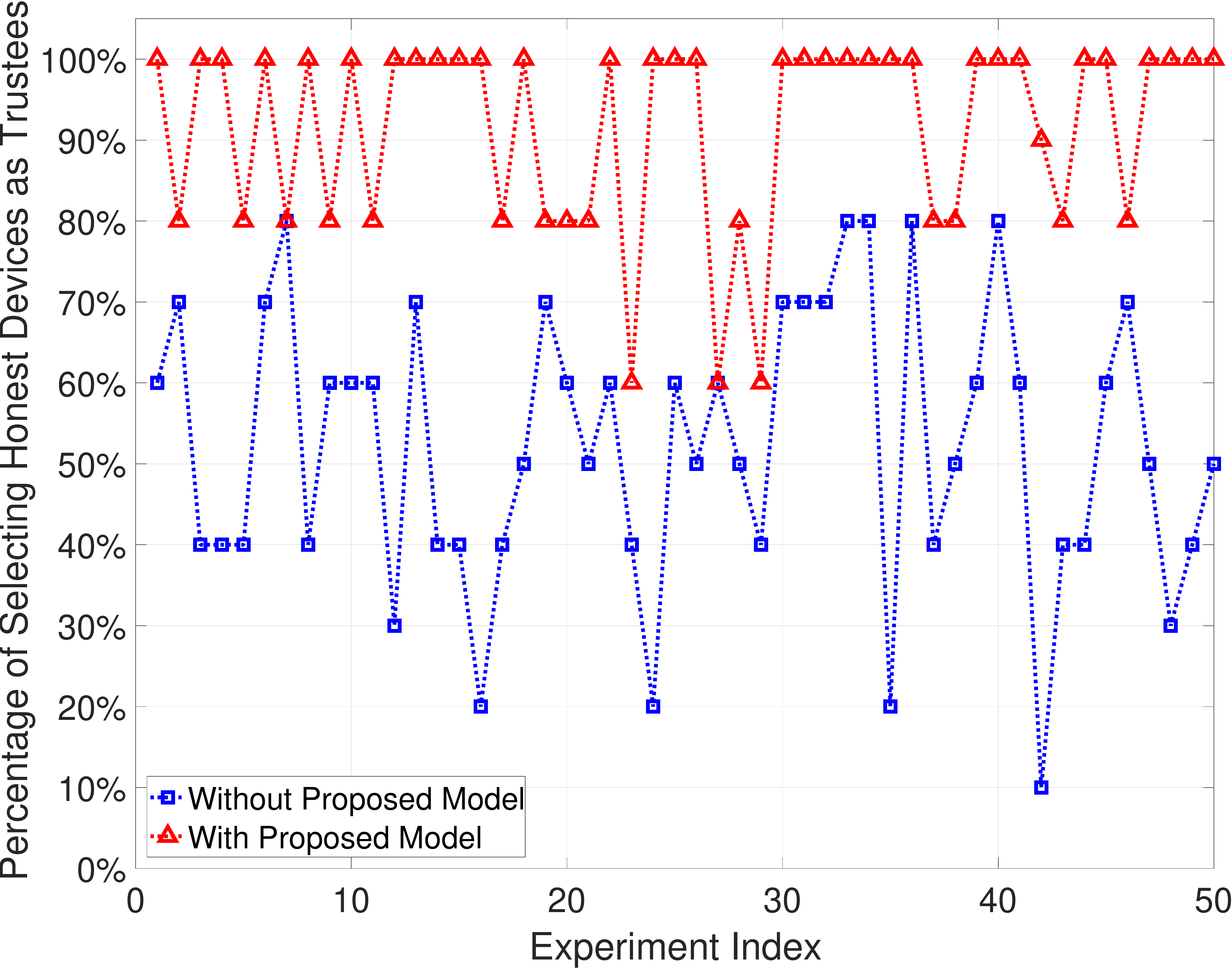}
\caption{Comparison of the percentages of honest devices.}
\label{fig:honestnodes}
\end{figure}

The experiment runs for 50 times, and the result is shown in Fig.~\ref{fig:honestnodes}.   The trustors choose the trustees with two different methods.   First, the trustors infer the trustworthiness of the trustees on the task with the analogous tasks the trustees performed previously.  This is the proposed trust model (marked as With Proposed Model in Fig.~\ref{fig:honestnodes}).  Second, the trustors deem the task as a completely new task and do not infer any trustworthiness values from previous experiences (marked as Without Proposed Model in Fig.~\ref{fig:honestnodes}).    At the end of an experiment run, each trustor sends a report message to the coordinator, which contains the identifier of the selected trustee.  A trustor may choose an honest or dishonest node device as its trustee.  The coordinator calculates the percentage of the trustors that have chosen the honest trustees.

As shown in the figure, the percentage of the trustors that have selected honest devices as the trustees for the task is higher when we use the proposed trust model.  The trustors select proper trustees with a high probability because they can reasonably infer trustworthiness of a trustee on performing a new task with the previous experiences.  If a trustee performed maliciously on a previous task, it is hard to gain sufficient trust to perform the analogous tasks.

\subsection{Transitivity in Trust Model}
\label{sec:transitivity}

In order to find out how the context affects the trust transitivity based on the proposed trust model, we simulate a scenario where there are multiple types of tasks in the network.  Each task consists of one or two characteristics.  Every network node keeps the trustworthiness records of two different tasks.  The characteristics of these tasks are randomly assigned in the simulation.  The total number of different characteristics of the tasks in the network is set to be 4, 5, 6, or 7.

Each trustor randomly generates a task delegation request.  With the conservative transitivity method, it sends out the delegation request to intermediate nodes who have recommended and potential trustees who have accomplished tasks that contain all the characteristics of the task.   When an intermediate node receives the delegation request, it relays the request to the proper trustees or the next intermediate nodes.   When a potential trustee receives the request, it responds.  With the aggressive transitivity method, the trustor sends out the delegation request to intermediate nodes who have recommended and potential trustees who have accomplished tasks that contain a part of the characteristics of the task.  When an intermediate node receives the delegation request, it relays the request to the proper trustees or the next intermediate nodes.  When a potential trustee receives the request, it waits for a preset period.   The same trustee may receive other delegation requests originated from the trustor.  If all the  characteristics of the task are covered in these requests, the trustee responds.  By this time, the trustee's capability of every characteristic of the task can be recommended to the trustor.   The trustor delegates the task to the trustee that has the highest trustworthiness value on this task.   Here, we only consider unilateral evaluation from the trustor toward the potential trustee in order not to mix the performances
of different features of the trust model.

For comparison, a traditional trust transfer method is also used.  In this method, the trustworthiness can only be inferred from a potential trustee that has accomplished the exact same task or transferred through an intermediate node that recommends the exact same task.   The trustor sends out delegation requests only to these potential trustees and intermediate nodes.   Therefore, the search is narrowed, and the trustor will find less potential trustees with the traditional method.  

For every task, a random number in $[0,1]$ is assigned to each network node to indicate its actual competence and willingness to accomplish the task.  If this task has two characteristics, this random number reveals the node's capability of handling each characteristic.   For a particular node in the social IoT, neighboring nodes that have direct experiences with it will establish the trustworthiness of this node that approaches its actual capability.  The trustworthiness will be transferred to other network nodes with the traditional, conservative transitivity, or aggressive transitivity method.

\begin{figure}[t]
\centering
\includegraphics[width=0.98\linewidth]{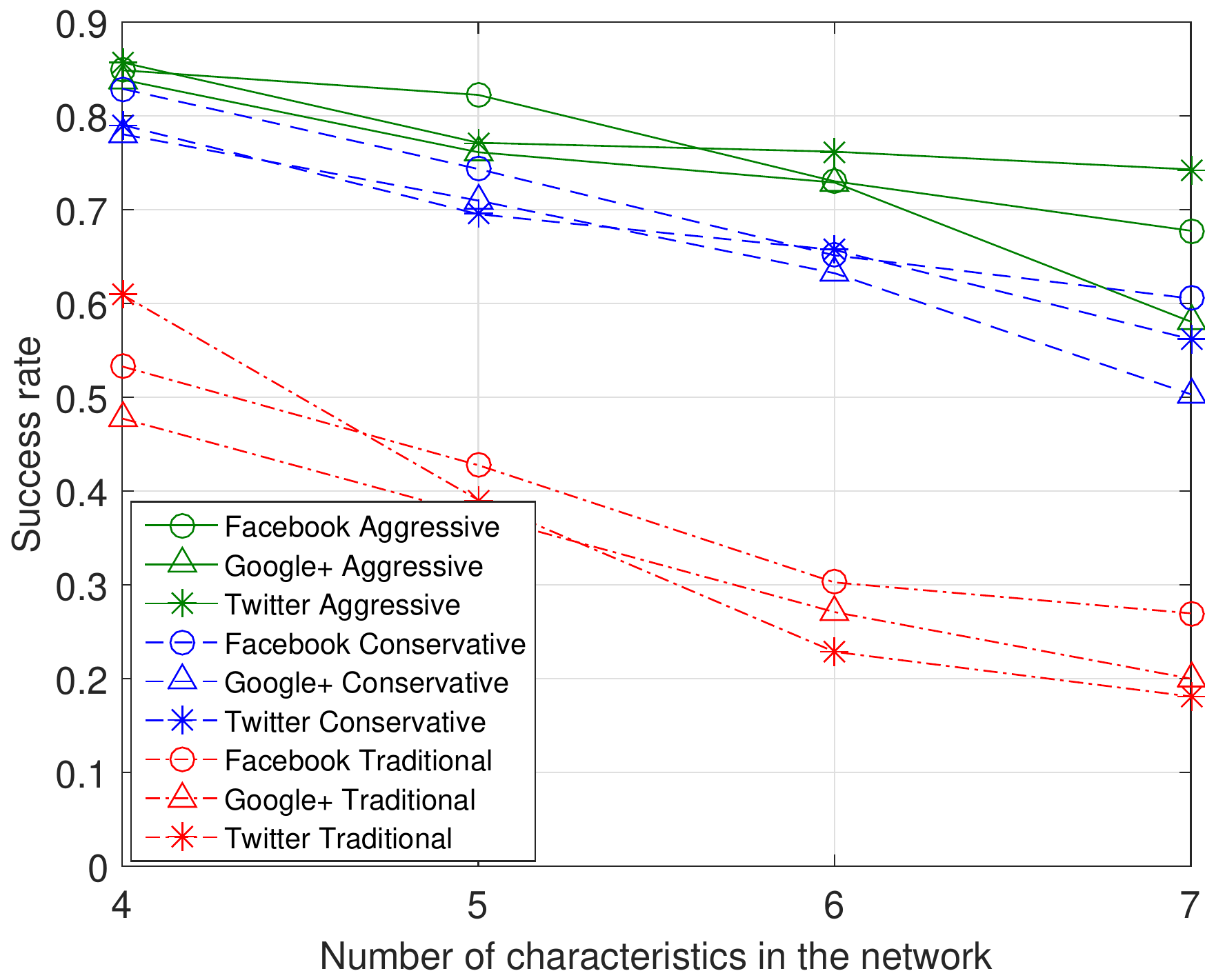}
\caption{Comparison of the success rates.}
\label{fig:section531}
\end{figure}

\begin{figure}[t]
\centering
\includegraphics[width=0.98\linewidth]{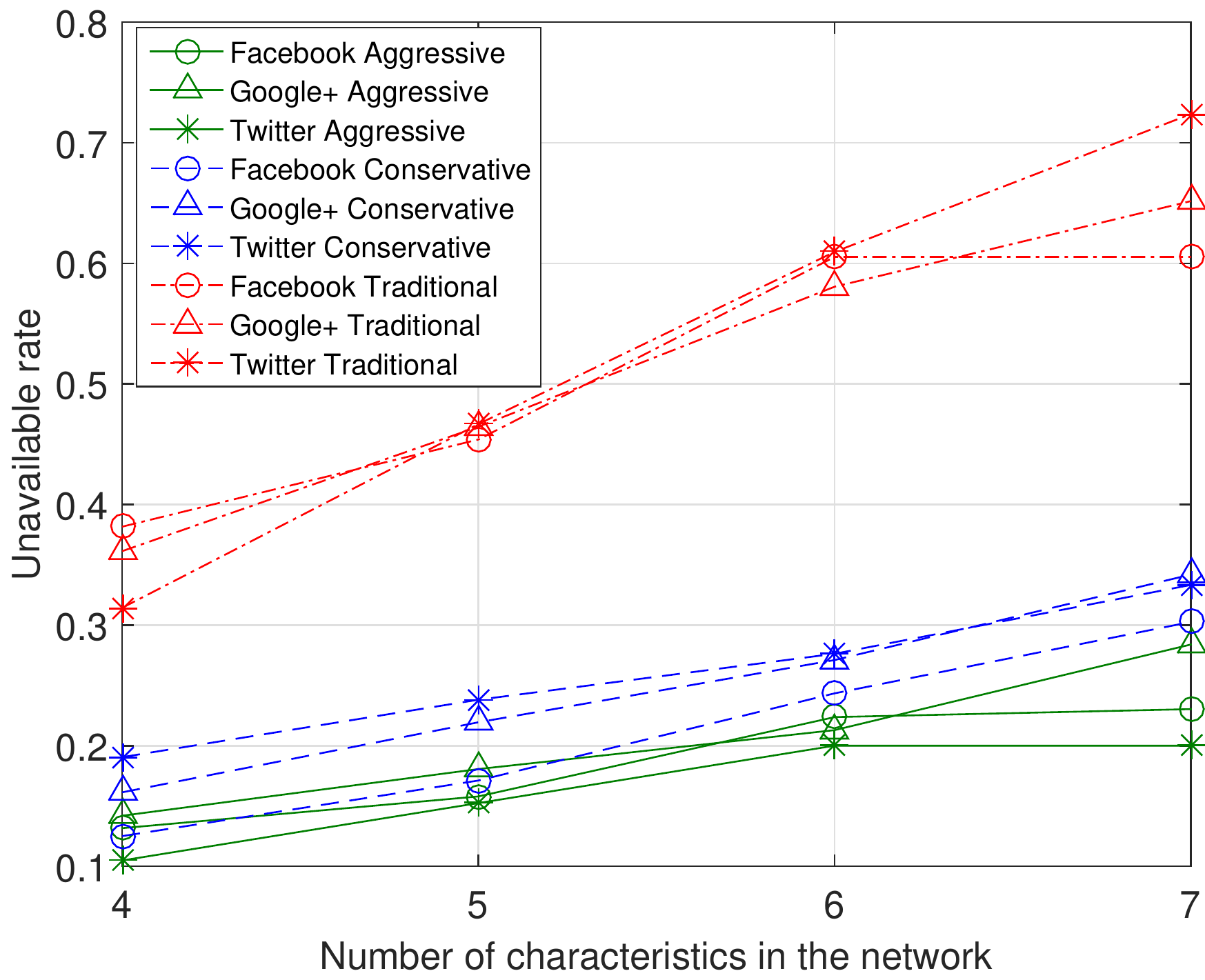}
\caption{Comparison of the unavailable rates.}
\label{fig:section532}
\end{figure}

\begin{figure}[t]
\centering
\includegraphics[width=0.98\linewidth]{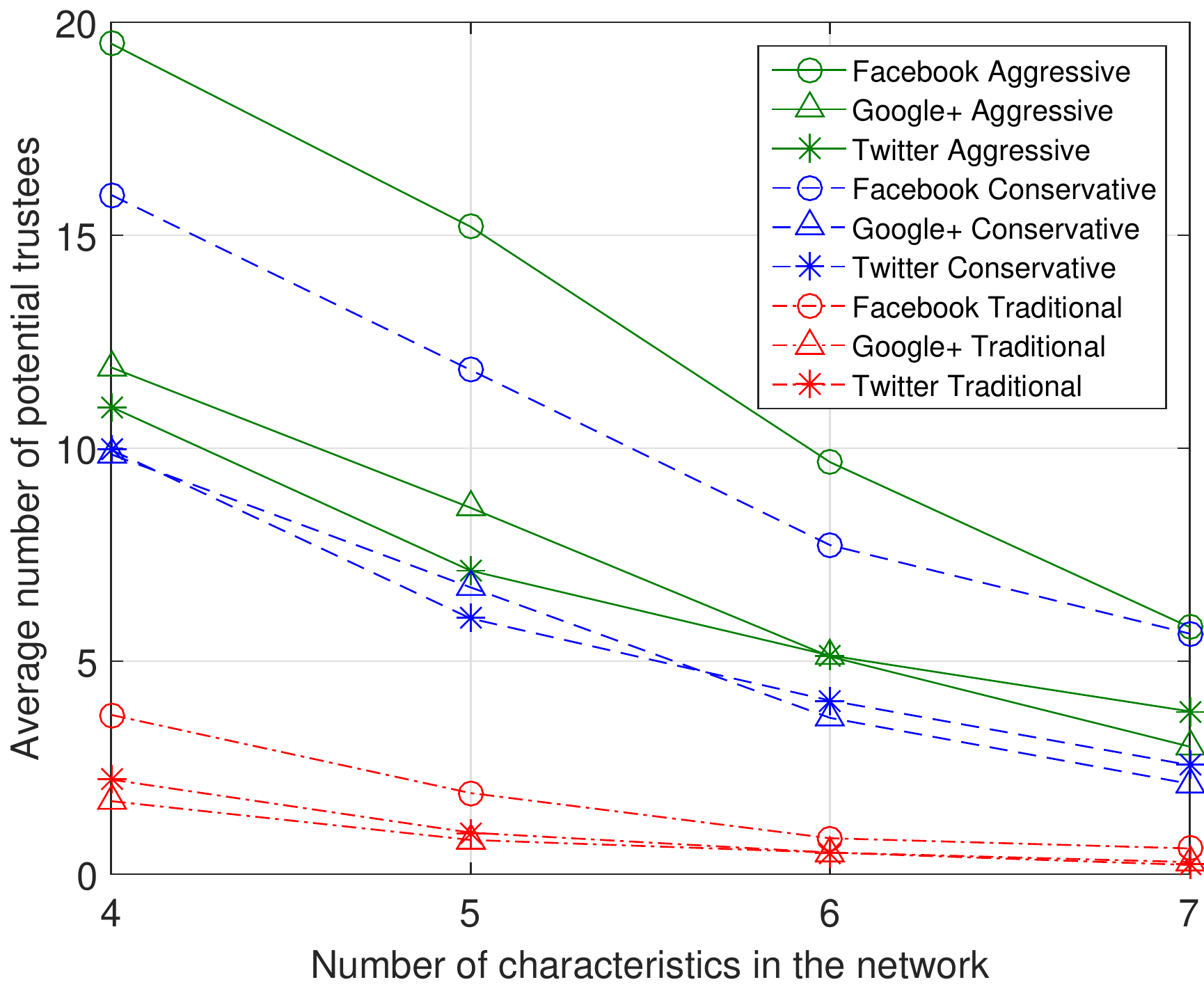}
\caption{Comparison of the average numbers of potential trustees.}
\label{fig:section533}
\end{figure}

For task delegations according to the trustworthiness values, Fig.~\ref{fig:section531} and Fig.~\ref{fig:section532} show the average success rates and the average unavailable rates with different trust transitivity methods over network models of Facebook, Google+, and Twitter.  The success rates decrease and the unavailable rates increase as the number of characteristics in the network increases.  This is because it is getting harder for the trustors to encounter the intermediate nodes and find the potential trustees with the same task context.

The task delegation processes with the methods of conservative trust transitivity and aggressive trust transitivity have better performance than the processes with the traditional trust transfer method. In Fig.~\ref{fig:section531}, the solid green and dash blue curves are above the dash-dot red curves.  Compared with the traditional method, the aggressive trust transitivity method has an improvement of more than 0.2 in success rate.  In Fig.~\ref{fig:section532}, the solid green and dash blue curves are below the dash-dot red curves.   The aggressive trust transitivity has an improvement of more than 0.3 in unavailable rate.
 This is because a trustor can find more potential trustees based on the two proposed trust transitivity methods. Meanwhile, the aggressive transitivity method slightly outperforms the conservative transitivity method because the former can guarantee that a trustor find even more potential trustees.  This trend is revealed in Fig.~\ref{fig:section533}.
The more potential trustees a trustor can find to delegate a task, the better chance that the task can be successfully accomplished.

To further evaluate the proposed models on the transitivity of trust, we use some real-world node properties of the three social networks to represent task characteristics.  The task delegation processes have results that are consistent with the random simulations.  The success rates, the unavailable rates, and the average numbers of potential trustees are summarized in Table~\ref{tab:ratecomparison}.
It shows that the conservative transitivity and aggressive transitivity methods outperform the traditional trust transfer method.  Take the Facebook subnetwork as an example, compared with the traditional method, the success rate increases to 57.89\% and 67.11\% from 27.63\% with conservative transitivity and aggressive transitivity, respectively.  And, the unavailable rate decreases to 37.50\% and 26.97\% from 66.45\% with conservative transitivity and aggressive transitivity, respectively.

\begin{table}[b]
\centering
\caption{Comparison of success rates, unavailable rates, and average numbers of potential trustees with real-world network node properties.}
\label{tab:ratecomparison}
\begin{tabular}{|r|c|c c c|}
\hline
 & Metric & Facebook & Google+ & Twitter \\
 \hline
  \hline
   & Success rate & 27.63\% & 28.39\% & 22.86\% \\ \cline{2-5}
  Trad. & Unavailable rate & 66.45\% & 60.00\% & 73.33\% \\ \cline{2-5}
   & Num.~potential trustees & 4.19 & 2.37 & 2.88\\
  \hline
 & Success rate & 57.89\% & 53.55\% & 48.57\% \\ \cline{2-5}
  Cons. & Unavailable rate & 37.50\% & 32.90\% & 45.71\% \\ \cline{2-5}
   & Num.~potential trustees & 10.63 & 5.92 & 5.99\\
  \hline
   & Success rate & 67.11\% & 59.35\% & 52.38\% \\ \cline{2-5}
  Aggr. & Unavailable rate & 26.97\% & 26.45\% & 35.24\% \\ \cline{2-5}
   & Num.~potential trustees & 11.60 & 6.53 & 6.35\\
  \hline
\end{tabular}
\end{table}

\begin{figure}[t]
\centering
\includegraphics[width=0.96\linewidth]{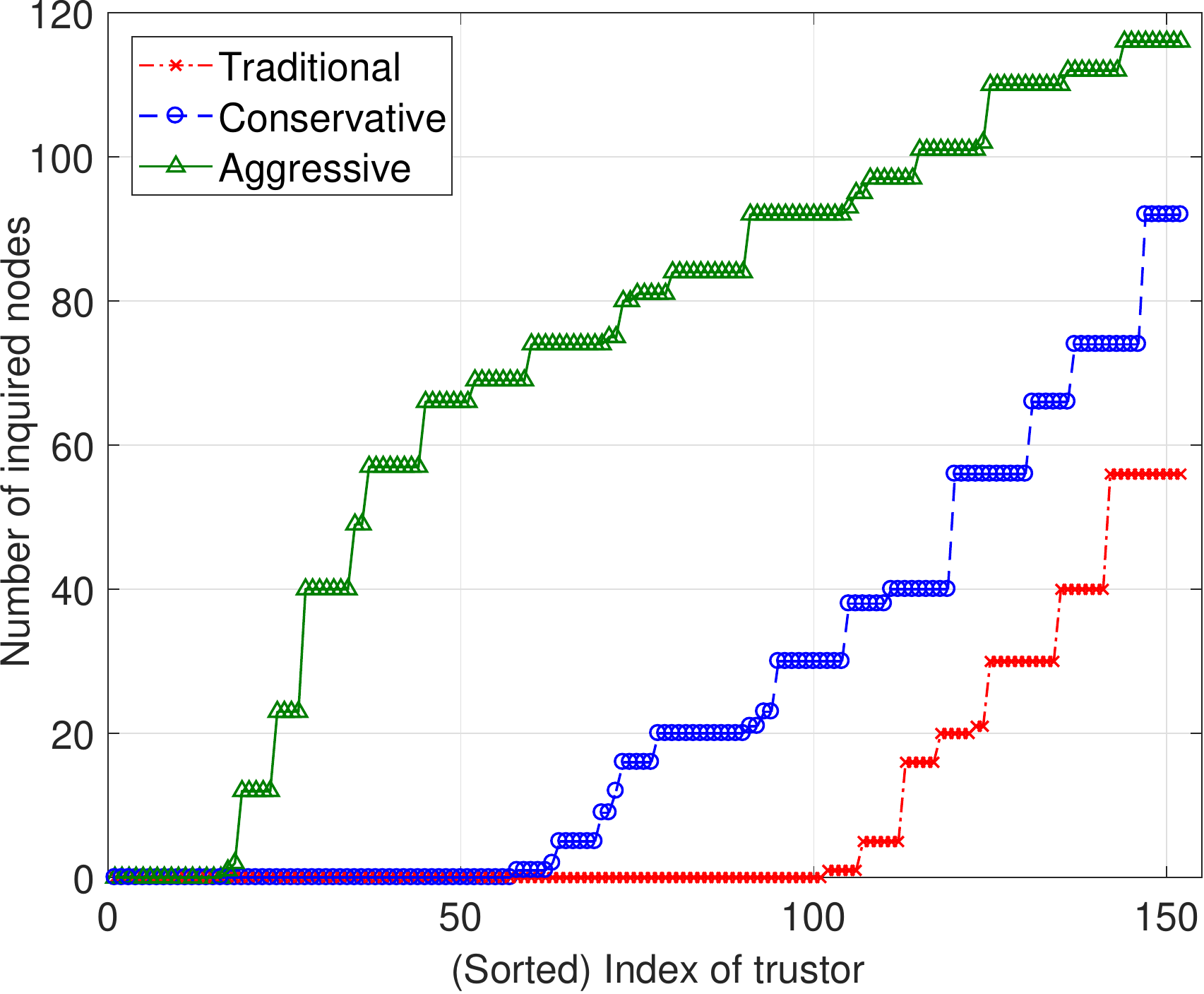}
\caption{Comparison of the numbers of inquired nodes with different trust transitivity methods.}
\label{fig:section534}
\end{figure}

The task delegation with aggressive transitivity of trust has the best performance.  However, it suffers from the largest search overhead.  The method with aggressive transitivity may involve network nodes that only have a portion of the characteristics of a task.  These nodes cannot accomplish the task themselves but work as intermediate nodes.  Extra effort is required to communicate with these nodes.  Fig.~\ref{fig:section534} illustrates the search overheads with different methods of trust transfer based on the Facebook subnetwork.  The search overhead is reflected in the number of network nodes that a trustor will interrogate to find its potential trustees.  The Aggressive curve represents the number of network nodes that the trustor communicates with, and each node has at least one related characteristic of the task.
The method of aggressive transitivity of trust increases the number of a trustor's potential trustees.  This is achieved with a cost of larger search overhead, i.e., the trustor interrogates more network nodes.  


\subsection{Trustworthiness with Delegation Results in Trust Model}
\label{sec:delegation}

In this simulation, we assign each potential trustee random values of the expected success rate, gain, damage, and cost.   The random values are in $[0,1]$.   The trustee behaves according to the success rate.  If the trustee accomplishes the task successfully, the trustor obtains the gain but pays the cost.  If the trustee fails to complete the task, the trustor suffers the damage and pays the cost. Every trustor selects its trustee among the potential trustees for task delegation with two strategies.  With the first strategy, the trustor only considers the success rates and delegates the task to the trustee with the highest success rate.  With the second strategy, as described in Section~\ref{sec:UpdatingTrustworthiness}, the trustor evaluates the potential trustees based not only on the success rate but also on the gain, damage, and cost. According to the task delegation results, the trustor updates the success rate, gain, damage, and cost that correspond to the task and that particular trustee node.

\begin{figure}[t]
\centering
\includegraphics[width=0.98\linewidth]{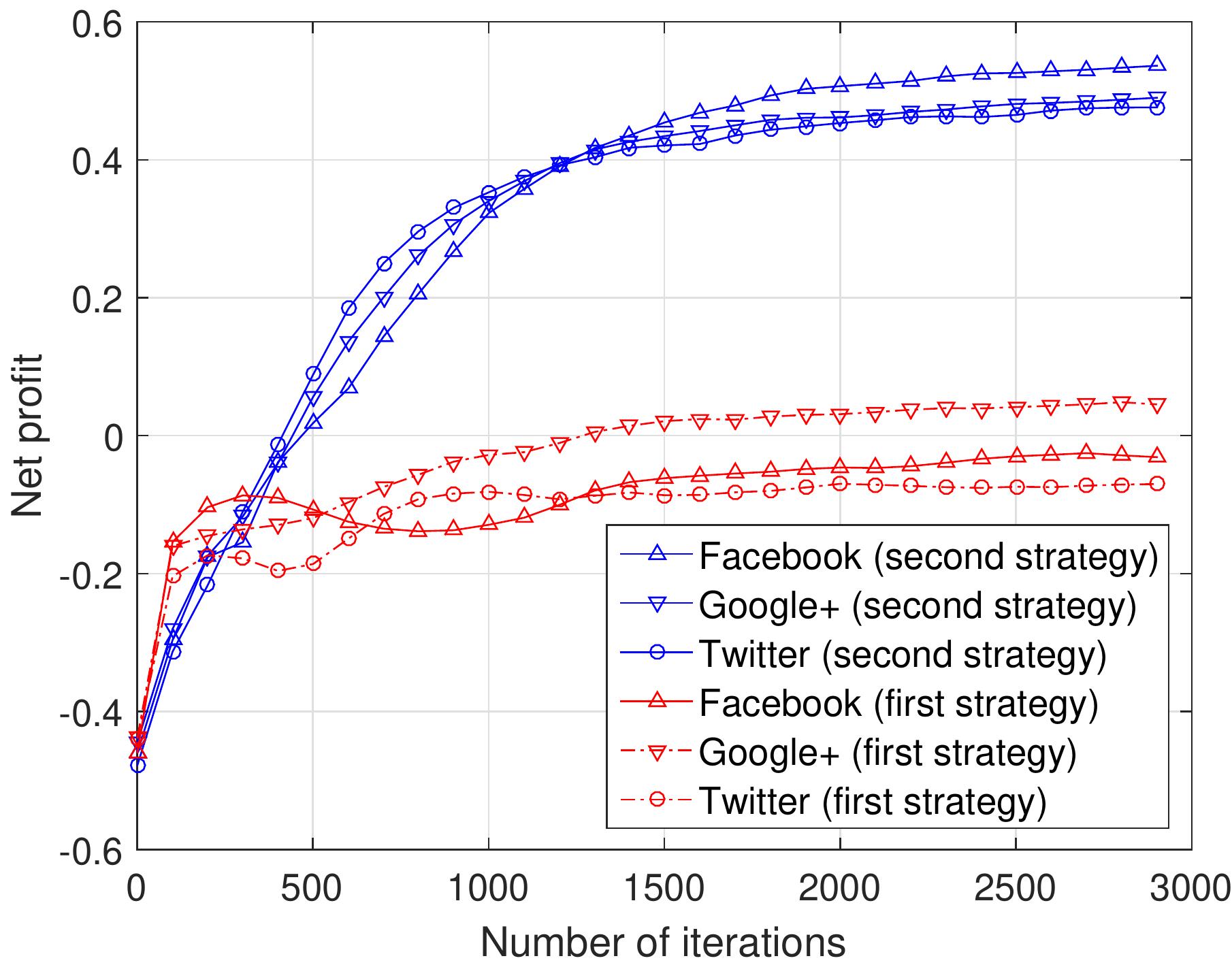}
\caption{Comparison of the net profits with iterative trustworthiness updates.}
\label{fig:section54}
\end{figure}

Fig.~\ref{fig:section54} shows the average net profits of task delegations in the subnetworks of Facebook, Google+, and Twitter.   With a forgetting factor $\beta = 0.1$, the success rate, gain, damage, and cost values are updated with iterations of continuous task delegations.   The average net profits are derived from the success rate, gain, damage, and cost and converge after many iterations.   For every subnetwork, a better net profit is obtained if the trustor evaluates the potential trustees with the second strategy, i.e., considering the success rate, gain, damage, and cost values.  With the first strategy, the simulation results in the subnetworks of Facebook and Twitter even show negative net profits.

\begin{figure}[t]
\centering
\includegraphics[width=0.98\linewidth]{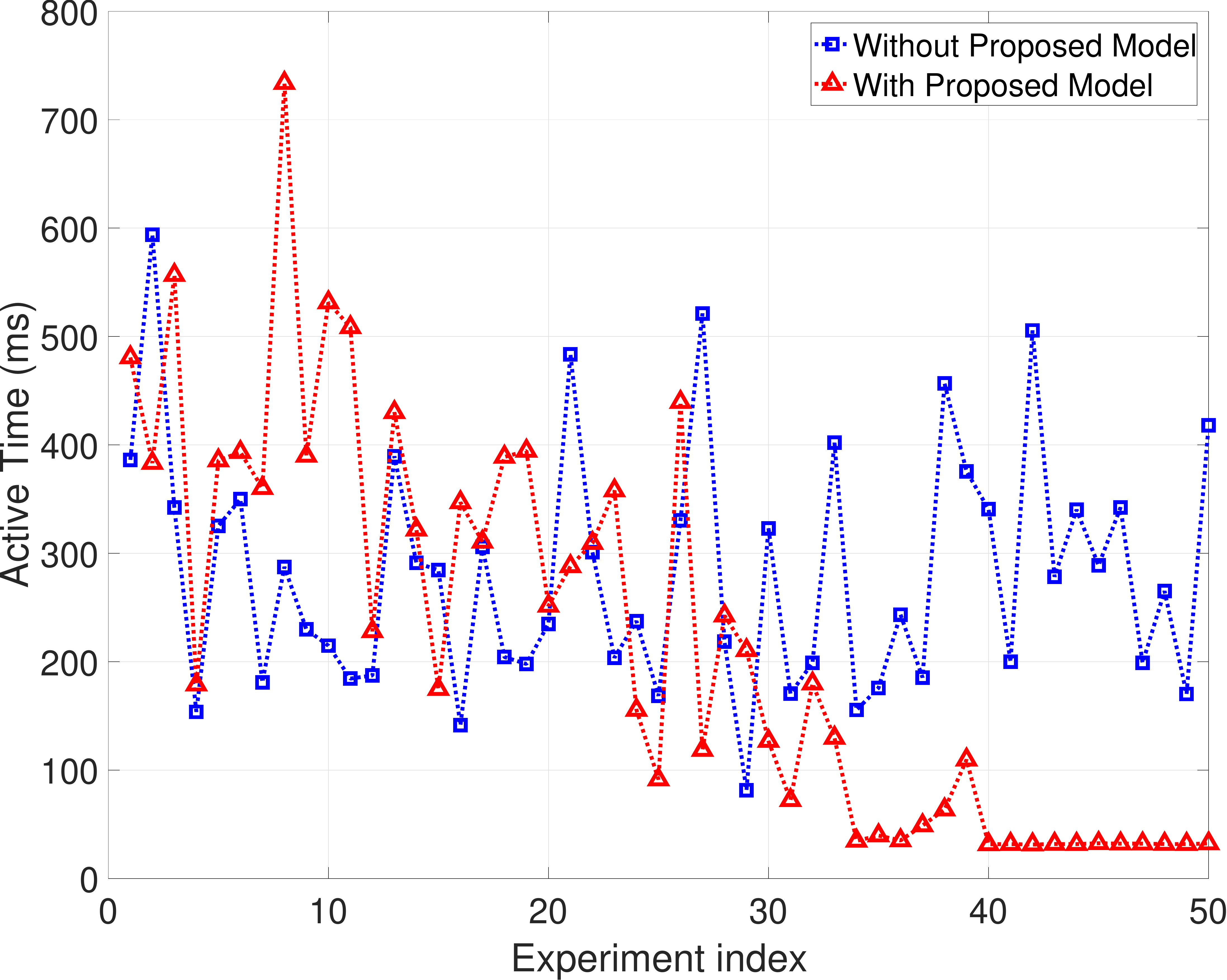}
\caption{Comparison of the active time.}
\label{fig:activetime}
\end{figure}

The experimental IoT network described in Section~\ref{sec:experiment_setup} is used to demonstrate how the proposed trust model deals with malicious behaviors.  
In the proposed model, the trustworthiness is evaluated with four different aspects, i.e., success rate, gain, damage, and cost.  It prevents the malicious nodes from promoting only a single aspect's value.   It is common that some IoT devices are sensitive to energy consumption.  These IoT devices try their best to conserve energy and extend the lifetime.   Usually, they stay in sleep mode and change to active mode only when it is necessary.   The energy consumption can be modeled as the cost.  

 In this experiment, the dishonest trustees send some fragment packages to prolong the interaction time with the trustors.   The trustors choose trustees with two different methods.  First, the trustors choose proper trustees based on both the gain and the cost (marked as With Proposed Model in Fig.~\ref{fig:activetime}).   Second, the trustors choose the trustees based only on the gain (marked as Without Proposed Model in Fig.~\ref{fig:activetime}).  
 
During the experiment, each trustor requests 50 tasks.   The average active time of the trustors is shown in Fig.~\ref{fig:activetime}.   As time goes on, the trustors using the proposed trust model can detect the malicious trustees, because the active time is much longer than usual.   As a result, the trustors do not choose those dishonest trustees anymore and the average active time is shortened.  Without the proposed model, however, the active time remains long over many tasks.  The experiment result reveals that the proposed trust model can be used to effectively identify malicious behaviors and thereby exclude malicious nodes.

\subsection{Trustworthiness with Dynamic Environment in Trust Model}
\label{sec:dynamic_environment}

\begin{figure}[t]
\centering
\includegraphics[width=0.98\linewidth]{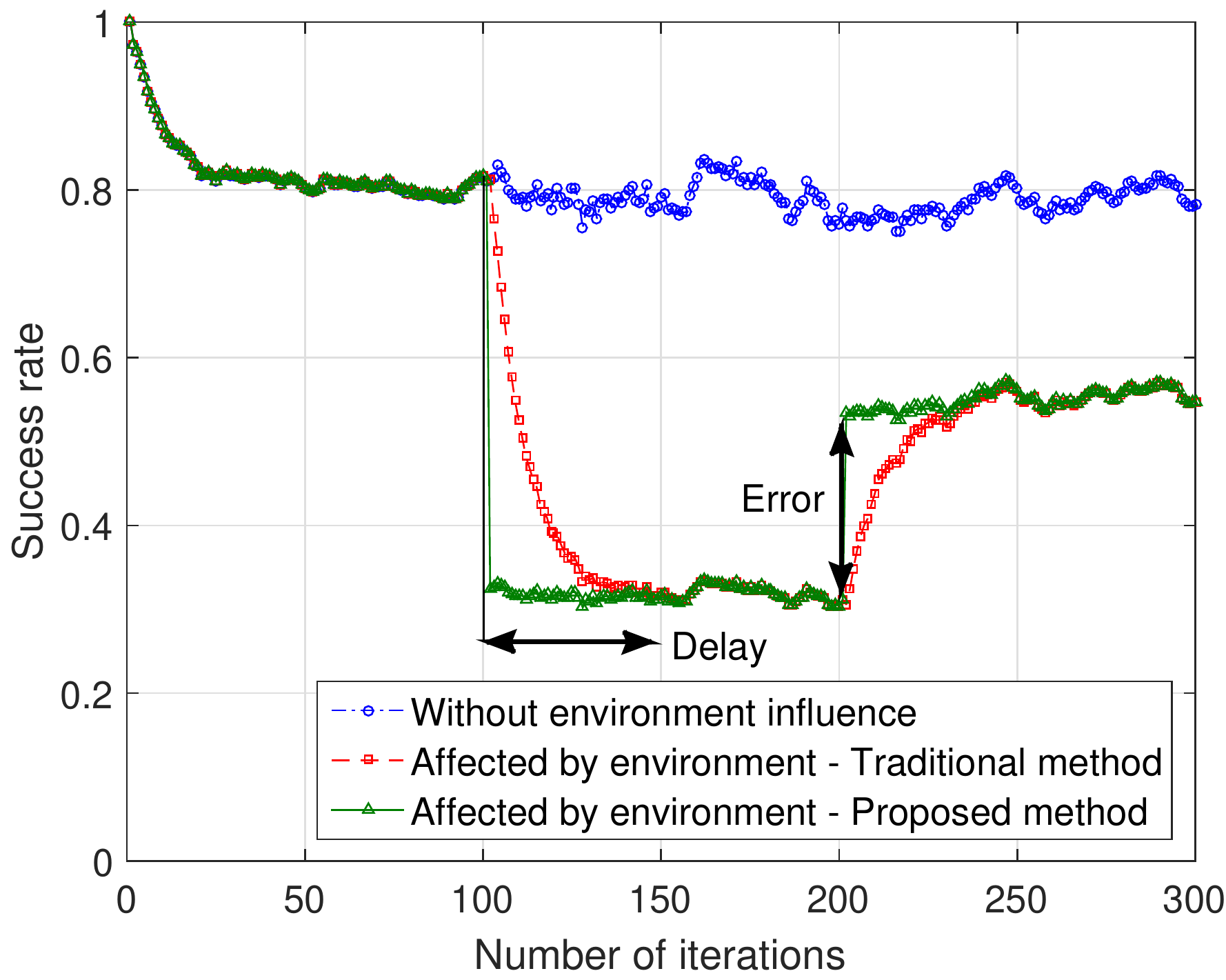}
\caption{Comparison of the success rates with non-ideal and changing environments.}
\label{fig:section55}
\end{figure}

We simulate an update process of the trustworthiness taking into account the influence of a dynamic environment.  Only the update of the success rate \eqref{eq:updateS} is used here to demonstrate how the changing environment affects the results.  For a random pair of trustor and trustee, Fig.~\ref{fig:section55} shows the success rates of delegating a particular task $\tau$ that are updated over iterations with a forgetting factor $\beta = 0.1$.  All the data points are averaged over 100 independent simulation runs.

The trustor initializes the expected success rate as 1.  A value of $S_{\mathrm{X}\leftarrow \mathrm{Y}}(\tau) = 0.8$ is assigned to the trustee to represent its actual competence and willingness to accomplish the task.  The trustor delegates the task to the trustee and updates the expected success rate according to the results.  During the first 100 update iterations, the environment is perfect and the instantaneous environments of the trustor and the trustee are equal to 1, i.e., $E_{\mathrm{X}} = E_{\mathrm{Y}} = 1$.  The expected success rates converge to 0.8.  During the second 100 update iterations, the environment deteriorates and $E_{\mathrm{X}} = E_{\mathrm{Y}} = 0.4$.  During the third 100 update iterations, the environment partially recovers such that $E_{\mathrm{X}} = E_{\mathrm{Y}} = 0.7$.

In the figure, the blue circles represent the expected success rates without the environment influence.  These rates converge to 0.8 which is the actual competence and willingness of the trustee for task $\tau$.  When the instantaneous environments change, the success rates change to $S_{\mathrm{X}\leftarrow \mathrm{Y}}(\tau) \cdot \min[E_{\mathrm{X}}, E_{\mathrm{Y}}] = 0.8 \times 0.4 = 0.32$ or $0.8 \times 0.7 = 0.56$.

The red squares show that, when the results are affected by the environment, how the expected success rates are updated according to the traditional method.  The traditional method does not distinguish the environment's variation from the task delegation results.  It takes quite some time for the traditional method to converge to the expected success rate when the environment suddenly changes.  Error and delay exist before convergence.

The green triangles show the expected success rates that are updated with a function $r(\cdot)$ as in \eqref{eq:functionr} to counter the environmental influence.  The expected success rates quickly track to the environment changes.  In general, the instantaneous environments, $E_{\mathrm{X}}$ and $E_{\mathrm{Y}}$, may not be difficult to obtained.  For example, these values reflect channel bandwidth, network workload, processing power, interference and noise, etc.  Nevertheless, it is relatively hard to construct the function $r(\cdot)$ that models how the environment affects the task delegation results.

\begin{figure}[t]
\centering
\includegraphics[width=0.98\linewidth]{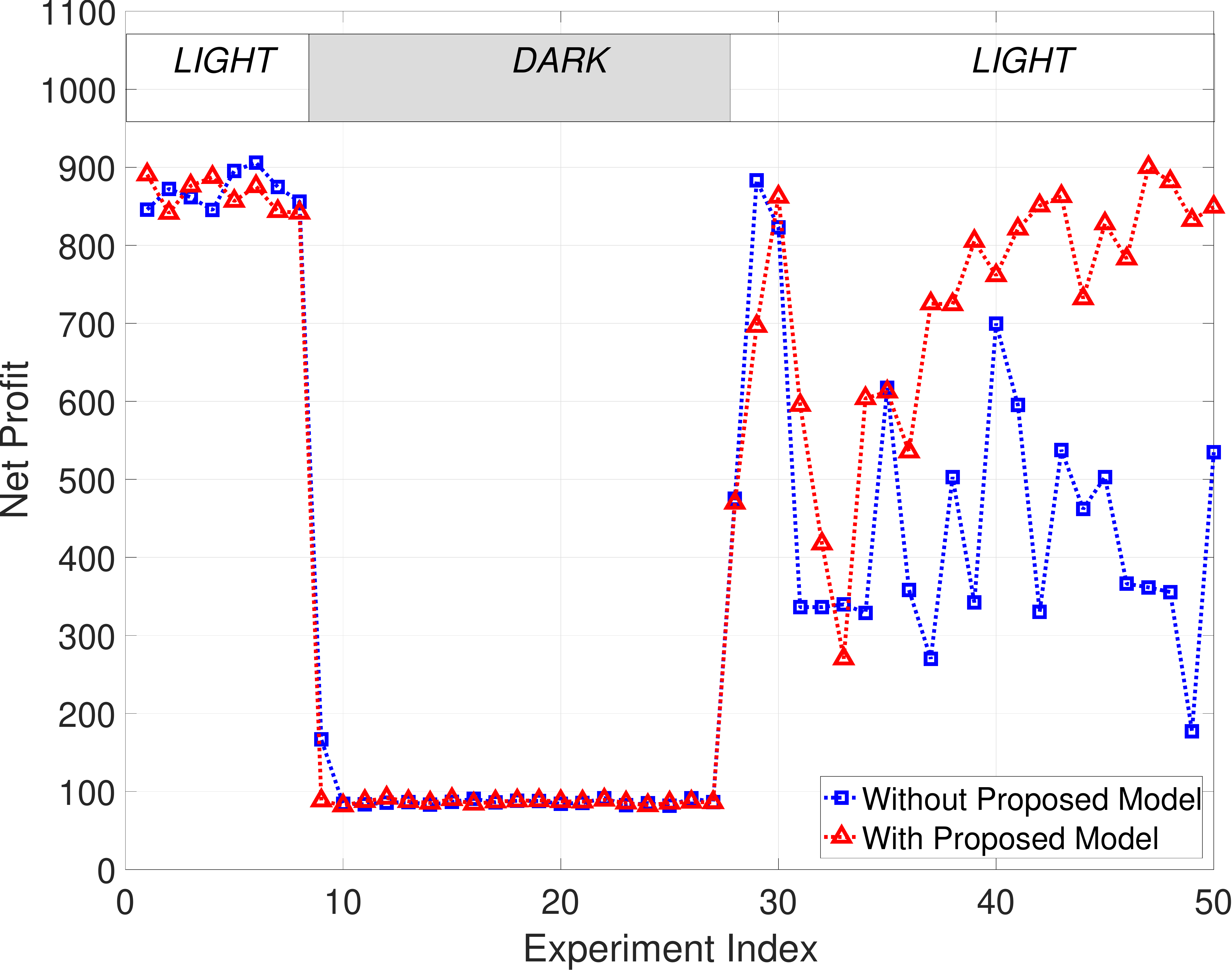}
\caption{Comparison of the net profits when the light condition changes and the dishonest trustees do not accept requests initially.}
\label{fig:netprofit1}
\end{figure}


In the experiment, we use the experimental IoT network described in Section~\ref{sec:experiment_setup}.   The node devices are installed with the Z-Stack and equipped with optical sensors.  The trust model with the dynamic environment factor is applied to distinguish the normal behaviors in a hostile environment from the malicious behaviors.

With the optical sensors, the performance of the trustee node is affected by the lighting condition.  For example, this can be the case in image acquisition.  In the experiment, there is a period of sufficient light followed by a dark period, and then it becomes light again.  The normal trustees serve the entire period and perform poorly in the dark situation.  The malicious trustees serve only during the last light period.  They behave maliciously from time to time, but the overall performance is better than that of the normal trustees performing in the dark.  With a forgetting factor $\beta = 0.1$, the trustors give the malicious trustees better evaluations, because the accumulated performance of the normal trustees is worse.

Fig.~\ref{fig:netprofit1} shows the net profit of the network with or without the proposed trust model.  With the proposed trust model, the trustors can remove the environment factor and appropriately evaluate the normal trustees during the dark period.   Over the last light period, more and more normal trustees are selected which replace the malicious trustees.  Therefore, the net profit returns to a high level.  Without the proposed trust model, the trustworthiness of the normal trustees declines due to the poor performance during the dark period.   They are not selected during the last light period, and the malicious trustees lower the net profit.  

\section{Conclusions and Discussions}
\label{sec:conclusion}
For the social IoT, a new model of trust among objects is proposed, and concepts of trust are clarified that overcome the limitations of the existing trust models.  The trust in the social IoT is depicted as a dynamic process.  It has a relational construct of six basic ingredients, i.e., (1) the trustor, (2) the trustee, (3) the goal, (4) the evaluation of trustworthiness, (5) the decision and its subsequent action and result, and (6) the context.  The distinctive features of the trust model are clarified in five aspects, i.e., (1) mutuality of trustor and trustee, (2) inferential transfer of trust with analogous tasks, (3) transitivity of trust, (4) trustworthiness updated with delegation results, and (5) trustworthiness affected by dynamic environment.  When the network interaction is based on the proposed trust model, the performance of the social IoT improves.  Simulations are conducted on the specific social IoT with network connectivity follows real-world social networks such as Facebook, Google+, and Twitter.  Experiments are carried out on an experimental IoT network.  The performance improvement is reflected in decreased abuse rates of task delegations with the trust mutuality model as well as increased success rates and decreased unavailable rates with the trust transitivity model.  The proposed methods of interaction also increase the net profits of the users and make network agents quickly adapt to a changing environment.

In the proposed trust model, the trustor and trustee perform the bilateral evaluation of the trustworthiness. Therefore, the vicious trustor cannot easily obtain services, and the malicious trustee cannot easily involve in trustors' tasks. Trustor and trustee evaluate each other on four different aspects, i.e., success rate, gain, damage, and cost, which can prevent the malicious nodes from promoting only a single aspect's value.  Also, this model is a characteristic-based model. Once the malicious trustee behaves maliciously in a task, it can affect subsequent evaluations of many other types of tasks which contain one or more the same characteristics. It means that the proposed model can detect malicious behavior effectively. Finally, the calculation of the trustworthiness in our model considers the dynamic environment, which can distinguish the normal behavior in a hostile environment from the malicious behavior.


\bibliographystyle{IEEEtran}
\bibliography{IEEEabrv,trust}

\begin{IEEEbiography}
[{\includegraphics[width=1in,height=1.25in,clip,keepaspectratio]{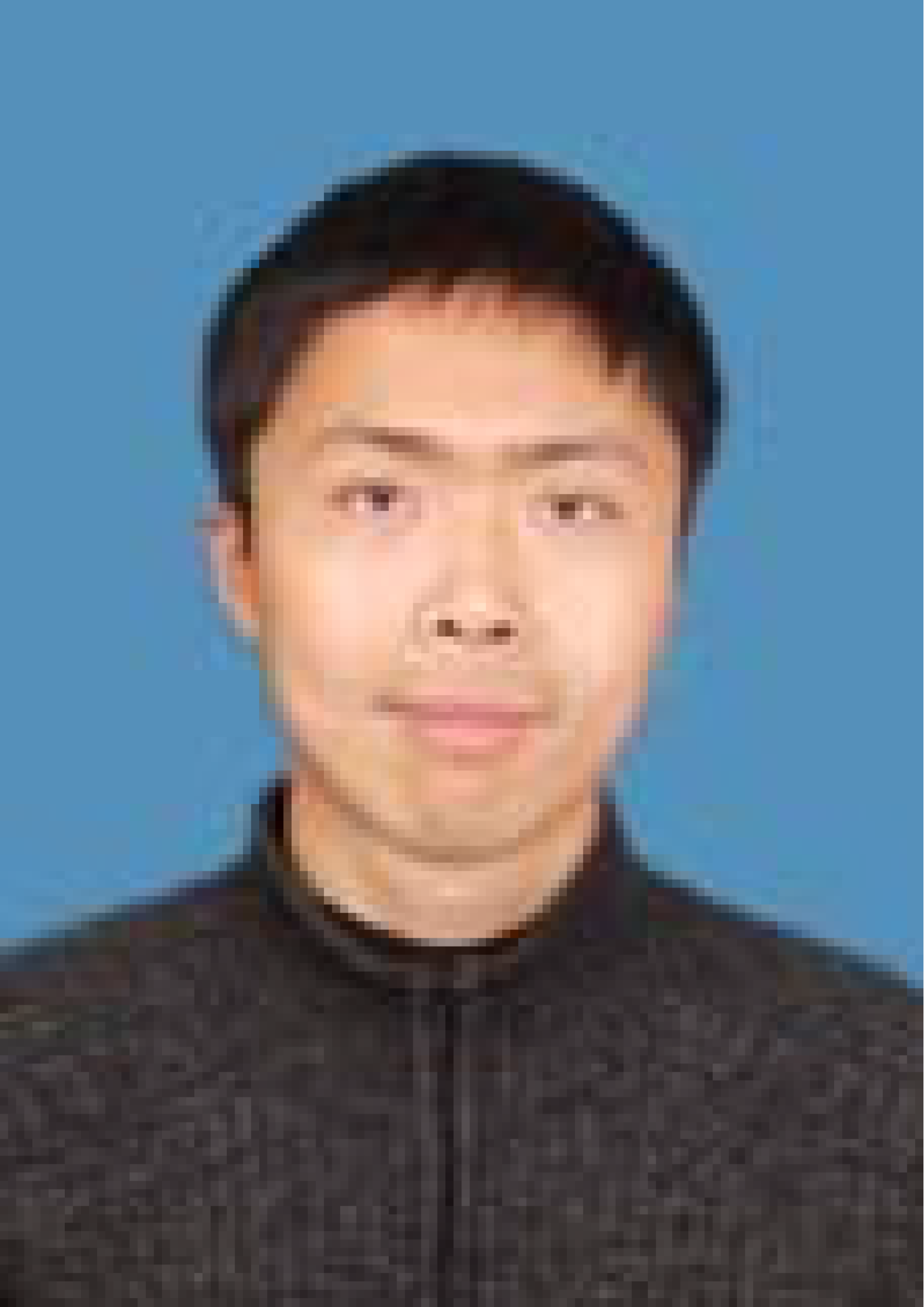}}]{Zhiting Lin}
received the BS and PhD degrees in electronics and information engineering from University of Science and Technology of China in 2004 and 2009, respectively.  From 2015 to 2016, he was a Visiting Scholar at the Department of Electrical and Computer Engineering of Baylor University, Waco, TX, USA. In 2011, he joined the Department of Electronics and Information Engineering, Anhui University, Hefei, Anhui, China.  He is currently an Associate Professor.  His research interests include Internet of Things and wireless social networks.  He has published about 40 papers and holds over 10 Chinese patents.
\end{IEEEbiography}

\begin{IEEEbiography}
[{\includegraphics[width=1in,height=1.25in,clip,keepaspectratio]{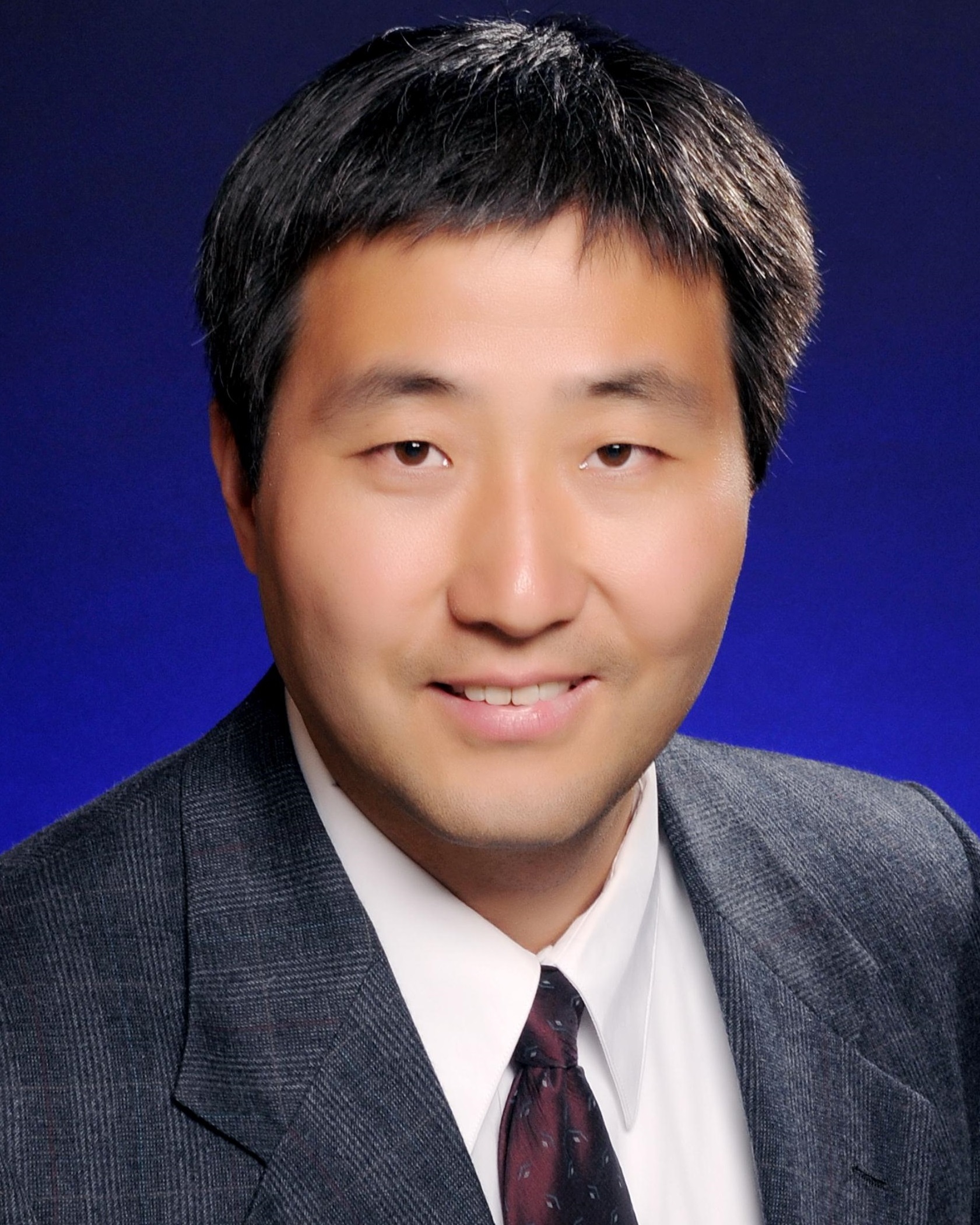}}]
{Liang Dong}
received the BS degree in applied physics with a minor in computer engineering from Shanghai Jiao Tong University, Shanghai, China, in 1996 and the MS and PhD degrees in electrical and computer engineering from The University of Texas at Austin, TX, USA, in 1998 and 2002, respectively. From 2002 to 2004, he was a Research Associate with the Department of Electrical Engineering, University of Notre Dame. From 2004 to 2011, he was an Assistant Professor then promoted to a tenured Associate Professor of the Department of Electrical and Computer Engineering, Western Michigan University. He joined the faculty of Baylor University, Waco, TX, in August 2011 as an Associate Professor of Electrical and Computer Engineering.  He held a visiting appointment at Stanford University in 2015.  His research interests include digital communications, digital signal processing, wireless communications and networking, and cyber-physical systems. He is a senior member of the IEEE.
\end{IEEEbiography}

\end{document}